\documentclass{CUP-JNL-DCE}

\usepackage{latexsym}
\usepackage{graphicx}
\usepackage{multicol,multirow}
\usepackage{amsmath,amssymb,amsfonts}
\usepackage{mathrsfs}
\usepackage{amsthm}
\usepackage{apacite}
\usepackage{rotating}
\usepackage{appendix}
\usepackage[authoryear]{natbib}
\usepackage{ifpdf}
\usepackage[T1]{fontenc}
\usepackage{times}
\usepackage{sourcesanspro}
\usepackage{newtxmath}
\usepackage{textcomp}%
\usepackage{xcolor}%
\usepackage{tikz}
\usepackage{etoolbox}
\usepackage[]{siunitx}
\usepackage{hyperref}
\usepackage{subcaption}

\usetikzlibrary{arrows,decorations.pathmorphing,backgrounds,positioning, fit}

\definecolor{echoreg}{RGB}{44, 177, 225}
\definecolor{olivegreen}{rgb}{0,0.6,0}
\definecolor{mymauve}{rgb}{0.58,0,0.82}
\definecolor{darkgreen}{RGB}{0,100,0}
\definecolor{steelblue}{RGB}{70,130,180}
\definecolor{darkorchid}{RGB}{153,50,204}

\newtoggle{redraw}
\newtoggle{redraw2}

\tikzset{%
pics/cube/.style args={#1/#2/#3/#4}{code={%
	\begin{scope}[line width=#4mm]
	\begin{scope}
	\clip (-#1,-#2,0) -- (#1,-#2,0) -- (#1,#2,0) -- (-#1,#2,0) -- cycle;
	\filldraw (-#1,-#2,0) -- (#1,-#2,0) -- (#1,#2,0) -- (-#1,#2,0) -- cycle;
	\end{scope}
\iftoggle{redraw}{%
}{%
	\begin{scope}
	\clip (-#1,-#2,0) -- (-#1-#3,-#2,-#3) -- (-#1-#3,#2,-#3) -- (-#1,#2,0) -- cycle;
	\filldraw (-#1,-#2,0) -- (-#1-#3,-#2,-#3) -- (-#1-#3,#2,-#3) -- (-#1,#2,0) -- cycle;
	\end{scope}
}
\iftoggle{redraw2}{%
}{
	\begin{scope}
	\clip (-#1,#2,0) -- (-#1-#3,#2,-#3) -- (#1-#3,#2,-#3) -- (#1,#2,0) -- cycle;
	\filldraw (-#1,#2,0) -- (-#1-#3,#2,-#3) -- (#1-#3,#2,-#3) -- (#1,#2,0) -- cycle;
	\end{scope}
}
	\node[inner sep=0] (-A) at (-#1-#3*0.5, 0, -#3*0.5) {};
	\node[inner sep=0] (-B) at (#1-#3*0.5, 0, -#3*0.5) {};
	
	\coordinate (-V) at (#1, #2);
	\coordinate (-W) at (#1, -#2);
	\end{scope}
}}}

\articletype{RESEARCH ARTICLE}
\jname{Data-Centric Engineering}
\jyear{2021}

\begin{document}

\begin{Frontmatter}

\title[ROMs applications with GP emulation]
{Multiphase flow applications of non-intrusive reduced-order models with Gaussian process emulation}

\author[1]{Themistoklis Botsas}
\author*[1,2,3]{Indranil Pan}\email{i.pan11@imperial.ac.uk}
\author[1,2]{Lachlan R. Mason}
\author[1,2]{Omar K. Matar}

\authormark{Botsas, Pan, Mason, Matar}

\address[1]{\orgname{The Alan Turing Institute}, \orgaddress{\street{96 Euston Rd}, \city{London}, \postcode{NW1 2DB}, \country{United Kingdom}}}

\address[2]{\orgname{Imperial College London}, \orgaddress{\street{Exhibition Rd, South Kensington}, \city{London}, \postcode{SW7 2BX}, \country{United Kingdom}}}

\address[3]{\orgname{Newcastle University}, \orgaddress{\city{Newcastle upon Tyne}, \postcode{NE1 7RU}, \country{United Kingdom}}}


\keywords{Reduced-order models; Autoencoders; Gaussian process; Deep Gaussian Process}


\abstract{Reduced-order models (ROMs) are computationally inexpensive simplifications of high-fidelity complex ones. Such models can be found in computational fluid dynamics where they can be used to predict the characteristics of multiphase flows. In previous work, we presented a ROM analysis framework that coupled compression techniques, such as autoencoders (AE), with Gaussian process (GP) regression in the latent space. This pairing has significant advantages over the standard encoding–decoding routine, such as the ability to interpolate or extrapolate in the initial conditions' space, which can provide predictions even when simulation data are not available. In this work, we focus on this major advantage and show its effectiveness by performing the pipeline on three multiphase flow applications. We also extend the methodology by using Deep Gaussian Processes (DGP) as the interpolation algorithm and compare the performance of our two variations, as well as another variation from the literature that uses Long short-term memory (LSTM) networks, for the interpolation.}


\begin{policy}[Impact Statement]
Reduced-order models are popular in various engineering fields since they replicate the behavior of their complex counterparts using minimal computational resources. By combining machine learning (ML) algorithms we can not only construct these models but also extend them in such a way that they incorporate knowledge from physical parameters, among other advantages. One advantage is that we can use these hybrid models to provide predictions from physical parameters even where data are not available, bypassing the standard expensive procedure of running new (physical and/or numerical) experiments. In the present study, we use one such combination in order to illustrate how this framework can be used in this manner and compare it with variations of other ML algorithms.

\end{policy}

\end{Frontmatter}

\section{Introduction}

Reduced-order models (ROMs) are widely applicable to various fields of science and engineering involving partial differential equations (PDEs), since they can speed up analyses and reduce computational requirements. ROMs are of particular interest to computational fluid dynamics (CFD), where they can be used to predict the characteristics of multiphase flows. For this work, we construct the main framework with CFD-related challenges in mind, such as computational complexity and need for physical parameter estimation.
Machine learning (ML) and Deep learning (DL) techniques are amongst the most popular choices employed in order to solve dimensionality reduction-related problems, including ROMs. Thus, algorithms such as autoencoders have been used in this manner \cite{kim2019deep} and are now considered an established and attractive choice.
In our previous work \citep{maulik2020latent}, we introduced a hybrid version of ROMs where we coupled three dimensionality reduction techniques, namely proper orthogonal decomposition (POD), convolutional autoencoders (CAE), and variational convolutional autoencoders (VAE) with interpolation in the latent space using Gaussian Process (GP) regression and focused on the various advantages of our methodology; these include uncertainty quantification (due to the deployment of GPs), the derivation of a finer temporal resolution, and enhanced interpretability.

In the present study, we focus on another major advantage of the methodology developed in our previous paper \cite{maulik2020latent}, which involves the ability to interpolate in parameter space. This is particularly significant for multiple engineering domains where small changes in parameter values can lead to significant differences in the temporal progression of the system, and running a simulator for all the required parameter values can be computationally intractable. 
A similar method that substitutes GPs for Long–short term memory recurrent neural networks (LSTMs) has recently emerged in \cite{maulik2021reduced}. We compare the performance of the GP- and LSTM-based interpolation techniques on the same data-sets and comment on the reasons underlying the differences observed. 
Finally, we extend the GP-based methodology by replacing the GPs with Deep Gaussian Processes (DGPs), an ML method that has gained traction in recent years and can be perceived as an extension of standard GPs in the same manner that a neural network is an extension of the generalised linear model.

%
The remainder of this paper is organised as follows. In the next section we present the general methodology and we briefly introduce all of the algorithms involved. In Section \ref{sec:Apps}, we demonstrate and compare the different variations of the methodology applied to three multiphase flow data-sets with increasing complexity. Finally, in Section \ref{sec:concl}, we summarise the main takeaways from our work and discuss the focus of our future research.

\section{Methodology}

The main pipeline that we will use for the remainder of this paper is similar to the one introduced in our previous paper \cite{maulik2020latent}. It combines two ML algorithms: a compression algorithm that takes a simulation in the form of time-related snapshots as input and outputs a latent space, and an interpolation algorithm, that is used as a regression model upon this space.

\subsection{Compression algorithms}\label{sec:compr}

\begin{figure}
    \centering{\includegraphics[width=\textwidth]{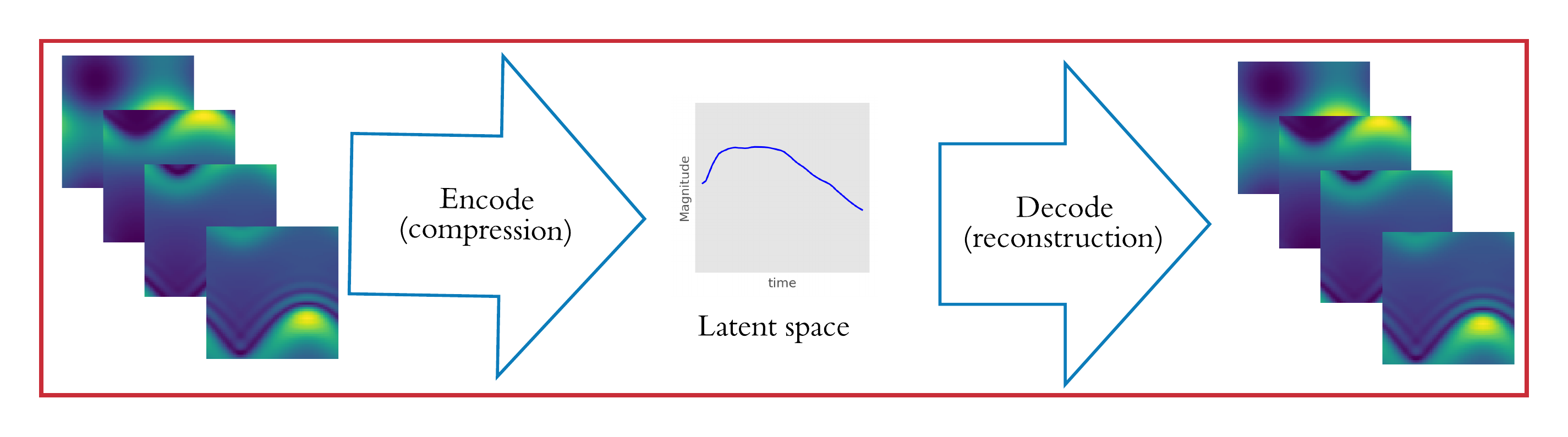}}
    \caption{Schematic of reduced-order modeling. Spatio-temporal outputs of a simulation (left) is being fed into an encoder and outputs a latent space (middle). The reconstruction of the original system (right) is the output of the decoder that uses as input the aforementioned latent space. Lower panel shows the interpolation and forecasting in the latent space for reconstruction}
    \label{schematic_vanilla}
\end{figure}

We focus on three different compression algorithms: proper orthogonal decomposition (POD), convolutional autoencoders (CAE), and variational convolutional autoencoders (VAE).
In the context of computational fluid dynamics, a simulation involves the numerical solution of a set of differential equations that usually requires considerable computational resources (particularly if the simulations are spatio-temporal and solved in 3D space). A set of simulations is fed into one of the aforementioned ML algorithms in the form of images (frames that correspond to simulation timestamps). The information from the simulations is compressed and summarised in the form of the latent space, where it can be further manipulated and analysed. Subsequently, the decompression portion of the algorithm can be used to reconstruct the original space. The whole process is shown in Figure~\ref{schematic_vanilla}.

\subsubsection{Proper Orthogonal decomposition}
The proper orthogonal decomposition (POD) \citep{berkooz1993proper} is a numerical method used to decompose a random vector field $\mathbf{u}({\mathbf{d}},t)$ (in which $\mathbf{d}$ and $t$ denote space, which in turn can be represented by an appropriate coordinate system, and time, respectively). Following the decomposition step, a new basis is created where the new variables are linear combinations of the originals such that the explainable system variance is maximised. The $n_r$-dimensional latent space is created by selecting the $n_r$ first components and discarding the rest.

To carry out a POD, we first take $p$ temporal snapshots of the field $\mathbf{s}$, which for $n$ spatial elements yields the matrix $\mathbf{S}$:
\[
\mathbf{S}=\begin{bmatrix}
u(d_1,t_1) & \cdots & u(d_n,t_1)  \\
\cdots & \cdots & \cdots  \\
u(d_1,t_p) & \cdots & u(d_n,t_p).  \\
\end{bmatrix}
\]
Then, we compute the covariance matrix $\mathbf{C}$ of $\mathbf{S}$ as:
\[
\mathbf{C} = \frac{1}{p-1}\mathbf{U}^T\mathbf{U},
\]
where $\mathbf{U}$ is an orthogonal matrix and consequently the eigenvalue diagonal matrix $\Lambda = \operatorname{diag}\{\lambda_1, \dots, \lambda_n\}$, where $\lambda_1 \leq \dots \leq \lambda_n$ and the corresponding eigenvector matrix $W$ are derived from:
\[
\mathbf{C}\mathbf{W}=\mathbf{W}\Lambda.
\]
Finally, the POD basis is given by:
\[
\mathbf{\theta} = \mathbf{U}\mathbf{W},
\]
and choosing the first $n_r$ columns of $\mathbf{\theta}$ results in the creation of an $n_r$-dimensional latent space.
The main advantage of the POD is its simplicity and ease of computation, but the quality of the results is not necessarily equivalent to that of more sophisticated methods.

\subsubsection{Convolutional autoencoders}

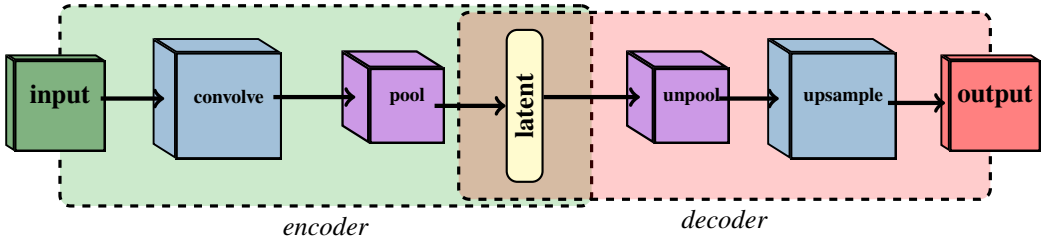
\begin{figure}
\centering
\begin{tikzpicture}

	\node (1) [draw, dashed, minimum height=7.5em, minimum width=20em, xshift=10em, fill=olivegreen, fill opacity=0.2, very thick, rectangle, rounded corners] {};
		\node (la1) [below=0em of 1] {{\emph{encoder}}};
	\node (2) [draw, dashed, minimum height=7em, fill = red, fill opacity=0.2,minimum width=20em, xshift=25em, very thick, rectangle, rounded corners] {};
		\node (la1) [below=0em of 2] {{\emph{decoder}}};

	\node[] (i2) {}; 
	\pic[fill=darkgreen!50] (I2) {cube={0.6/0.6/0.2/0.5}};
	\node [above =0mm of i2, anchor=center] (i2t){${\bf input}$};
	
	\togglefalse{redraw}
	\togglefalse{redraw2}
	
  	\node[right=6em of i2] (y) {};
  	\pic[right=6em of i2, fill=steelblue!50] (Y) {cube={0.7/0.7/0.5/0.5}};
  	\node [above left=0mm of y, anchor=center] (yt){\footnotesize{${\bf convolve}$}};

  	\node[right=6em of y] (y1) {};
  	\pic[right=6em of y, fill=darkorchid!50] (Y1) {cube={0.5/0.5/0.6/0.6}};
  	\node [above left=0mm of y1, anchor=center] (y1t){\footnotesize{${\bf pool}$}};

    \node[right=3em of y1] (rect) [draw,thick,minimum width=0.4cm,minimum height=2cm, fill=yellow!23, rounded corners] {\rotatebox{90}{${\bf latent}$}};

    \node[right=10em of y1] (y2) {};
  	\pic[right=10em of y1, fill=darkorchid!50] (Y2) {cube={0.5/0.5/0.6/0.6}};
  	\node [above left=0mm of y2, anchor=center] (y2t){\footnotesize{${\bf unpool}$}};
  	
  	\node[right=5em of y2] (y3) {};
  	\pic[right=5em of y2, fill=steelblue!50] (Y3) {cube={0.7/0.7/0.5/0.5}};
  	\node [above left=0mm of y3, anchor=center] (y3t){\footnotesize{${\bf upsample}$}};
  	
  	\node[right=5em of y3] (z2) {};
  	\pic[right=5em of y3, fill=red!50] (Z2) {cube={0.6/0.6/0.2/0.5}};
  	\node [above left=0mm of z2, anchor=center] (z2t){${\bf output}$};
  	
  	\draw [->, ultra thick] (I2-B|-Y-A) -- node[above] {} (Y-A);
  	
  	\draw [->, ultra thick] (Y-B|-Y1-A) -- node[above] {} (Y1-A);
  	
  	\draw [->, ultra thick] (Y1-B|-rect.west) -- node[above] {} (rect.west);
  	
  	\draw [->, ultra thick] (rect.east|-Y2-A) -- node[above] {} (Y2-A);
  	
  	\draw [->, ultra thick] (Y2-B|-Y3-A) -- node[above] {} (Y3-A);
  	
  	\draw [->, ultra thick] (Y3-B|-Z2-A) -- node[above] {} (Z2-A);
  	
  	\color{black}
  	
  	\toggletrue{redraw}
  	\toggletrue{redraw2}

  	\togglefalse{redraw2}
  	\toggletrue{redraw2}

\end{tikzpicture}
\caption{Architecture of a convolutional autoencoder (CAE). The input frames are fed into the \textit{encoder}, where convolutional (convolve) and pooling (pool) layers are used, in order to produce the \textit{latent space}. The reverse scheme is used for the \textit{decoder}, where unpooling (unpool) and upsampling (upsample) layers are used to produce an output as similar as possible to the original input} \label{fig:CAE_archit}
\end{figure}

Autoencoders are classes of neural network algorithms used primarily for unsupervised learning purposes, such as dimensionality reduction, data generation, and feature learning. Their structure is based on a bottleneck that combines two individual components: the \emph{encoder}, which is a neural network that passes the input through layers that consist of a decreasing number of neurons, up until the bottleneck, where it outputs the \emph{latent space}; and the \emph{decoder}, which has the opposite architecture, uses the latent space as input and aims to reconstruct the original data by minimising the reconstruction error.
For this work, we use convolutional autoencoders (CAE), which is a class of autoencoders that includes layers with convolutions \citep{lecun1995convolutional}, i.e. a set of filters that extract specific features from images. The output $y_{ijk}$ of a typical neuron in a convolutional layer has the form:
\[
y_{ijk} = \varphi (\mathbf{f}_i \ast \mathbf{p}_{jk} + b_{i}),
\]
where $\varphi$ is an activation function, $\mathbf{f}_i$ is a single filter, $\mathbf{p}_{jk}$ is a patch of data that shifts according to the dimensions $j$ and $k$, and $b_{i}$ is a bias term. In practice, the convolutional layers learn different features and patterns from the original data, 
particularly useful in image processing.

Another type of layer found in a convolutional neural network (CNN) is a pooling layer, which generally follows one (or more than one) convolutional layer with the aim of sub-sampling and summarising the information from the filters. In a typical CNN, the continuous alternation of convolutional and pooling layers is how a neural network can extract high- and low-level features.
%
In CAEs, the decoder comprises the opposite structure to that of the encoder. Instead of the convolutional and pooling layers, it consists of upsampling and unpooling layers respectively, where it produces an output of the same size as the input data using nearest-neighbour interpolation. A loss function such as the mean squared error (MSE) is then used during training in order to update the neural network's weights in a manner that minimises the reconstruction error between the input and the output.
The general structure of the CAE with all the components described above is shown in Figure \ref{fig:CAE_archit}.

\subsubsection{Variational convolutional autoencoders}

\begin{figure}
\centering
\begin{tikzpicture}

	\node (1) [draw, dashed, minimum height=7.5em, minimum width=20em, xshift=10em, fill=olivegreen, fill opacity=0.2, very thick, rectangle, rounded corners] {};
		\node (la1) [below=0em of 1] {{\emph{encoder}}};
	\node (2) [draw, dashed, minimum height=7em, fill = red, fill opacity=0.2,minimum width=20em, xshift=25em, very thick, rectangle, rounded corners] {};
		\node (la1) [below=0em of 2] {{\emph{decoder}}};

	\node[] (i2) {}; 
	\pic[fill=darkgreen!50] (I2) {cube={0.6/0.6/0.2/0.5}};
	
	\togglefalse{redraw}
	\togglefalse{redraw2}
	
  	\node[right=6em of i2] (y) {};
  	\pic[right=6em of i2, fill=steelblue!50] (Y) {cube={0.7/0.7/0.5/0.5}};
  	\node [above =0mm of i2, anchor=center] (i2t){${\bf input}$};
  	\node [above left=0mm of y, anchor=center] (yt){\footnotesize{${\bf convolve}$}};
  	
  	\node[right=6em of y] (y1) {};
  	\pic[right=6em of y, fill=darkorchid!50] (Y1) {cube={0.5/0.5/0.6/0.6}};
  	\node [above left=0mm of y1, anchor=center] (y1t){\footnotesize{${\bf pool}$}};
  	
    \node[above right=2.5em of y1] (rectmu) [draw,thick,minimum width=0.4cm,minimum height=0.5cm, fill=yellow!23, rounded corners] {${\bf \mu}$};
    
    \node[below right=2.5em of y1] (rectsigma) [draw,thick,minimum width=0.4cm,minimum height=0.5cm, fill=yellow!23, rounded corners] {${\bf \sigma}$};

    \node[right=3.6em of y1] (rect) [draw,thick,minimum width=0.4cm,minimum height=2cm, fill=yellow!23, rounded corners] {\rotatebox{90}{${\bf latent}$}};

    \node[right=10em of y1] (y2) {};
  	\pic[right=10em of y1, fill=darkorchid!50] (Y2) {cube={0.5/0.5/0.6/0.6}};
  	\node [above left=0mm of y2, anchor=center] (y2t){\footnotesize{${\bf unpool}$}};
  	
  	\node[right=5em of y2] (y3) {};
  	\pic[right=5em of y2, fill=steelblue!50] (Y3) {cube={0.7/0.7/0.5/0.5}};
  	\node [above left=0mm of y3, anchor=center] (y3t){\footnotesize{${\bf upsample}$}};
  	
  	\node[right=5em of y3] (z2) {};
  	\pic[right=5em of y3, fill=red!50] (Z2) {cube={0.6/0.6/0.2/0.5}};
  	\node [above left=0mm of z2, anchor=center] (z2t){${\bf output}$};
  	
  	\draw [->, ultra thick] (I2-B|-Y-A) -- node[above] {} (Y-A);
  	
  	\draw [->, ultra thick] (Y-B|-Y1-A) -- node[above] {} (Y1-A);
  	
  	\draw [->, ultra thick] (Y1-B) -- node[above] {} (rectmu.west);
  	
  	\draw [->, ultra thick] (Y1-B) -- node[above] {} (rectsigma.west);
  	
  	\draw [->, ultra thick] (rectmu.east) -- node[above] {} (rect.west);
  	
  	\draw [->, ultra thick] (rectsigma.east) -- node[above] {} (rect.west);
  	
  	\draw [->, ultra thick] (rect.east|-Y2-A) -- node[above] {} (Y2-A);
  	
  	\draw [->, ultra thick] (Y2-B|-Y3-A) -- node[above] {} (Y3-A);
  	
  	\draw [->, ultra thick] (Y3-B|-Z2-A) -- node[above] {} (Z2-A);
  	
  	\color{black}
  	
  	\toggletrue{redraw}
  	\toggletrue{redraw2}

  	\togglefalse{redraw2}
  	\toggletrue{redraw2}

\end{tikzpicture}
\caption{Architecture of a variational convolutional autoencoder. It is similar to the CAE equivalent, with the additional assumption that the latent space is a set of multivariate Gaussian distributions, and can be described as $z \sim N(\mu,\sigma^2)$} \label{fig:VAE_archit}
\end{figure}
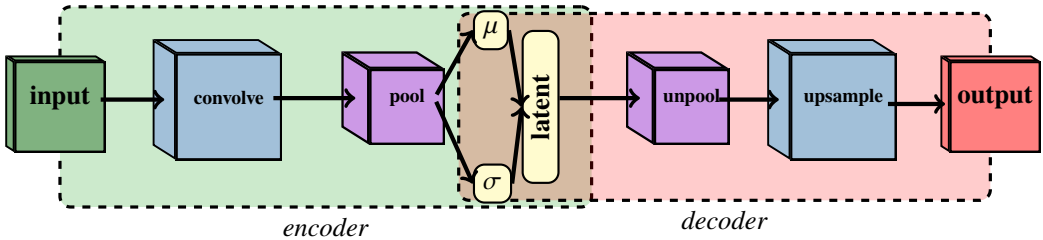

A similar class of algorithms is variational convolutional autoencoders (VAEs) \citep{kingma2013auto}. The main difference between these and their conventional convolutional counterparts is that the latent space is provided in the form of a probability distribution, usually a Gaussian. In practice this representation acts as \emph{regularisation} for the latent space, which can be particularly beneficial when quality of the data is poor. In order to reconstruct the input, a sample is drawn from the latent space distribution.

The encoder and decoder of a VAE can be described as the functions $q(\mathbf{z}|\mathbf{x})$ and $p(\mathbf{x}^{\prime}|\mathbf{z})$ respectively, where $\mathbf{x}$ is the input, $\mathbf{z}$ is the latent space and $\mathbf{x}^{\prime}$ is the output. The latent space follows a Gaussian distribution $\mathbf{z} \sim \mathcal{N}(\mathbf{\mu}, \mathbf{\sigma}^2)$ due to the Kullback–Leibler divergence (KL divergence) $D_{\mathrm{KL}}(q(\mathbf{z}|\mathbf{x}_i)||p(\mathbf{z}|\mathbf{x}_i))$.
The VAE loss has two components: the reconstruction loss (similar to the CAE) and the KL loss,
\[
E_{q(\mathbf{z}|\mathbf{x}_i)}[\log{p(\mathbf{x}_i|\mathbf{z})}]- D_{\mathrm{KL}}(q(\mathbf{z}|\mathbf{x}_i)||p(\mathbf{z}|\mathbf{x}_i)),
\]
which ensure that the output is as similar as possible to the input and that the distribution of $z$ is Gaussian, respectively.
The practical difference between a CAE and a VAE is that, as shown in Figure \ref{fig:VAE_archit}, the output of the VAE encoder consists of two vectors $\mu$ and $\sigma$. Given $\epsilon \sim N(0,I)$, where $I$ is the identity matrix we can use the \emph{reparameterisation trick} \citep{kingma2013auto} and write the latent space as $z = \mu + \sigma\epsilon$, which is a form suitable for training. 
For the implementation of the autoencoders we use the TensorFlow package \citep{abadi2016tensorflow}.

\subsection{Interpolation algorithms}

\begin{figure}
    \centering{\includegraphics[width=\textwidth]{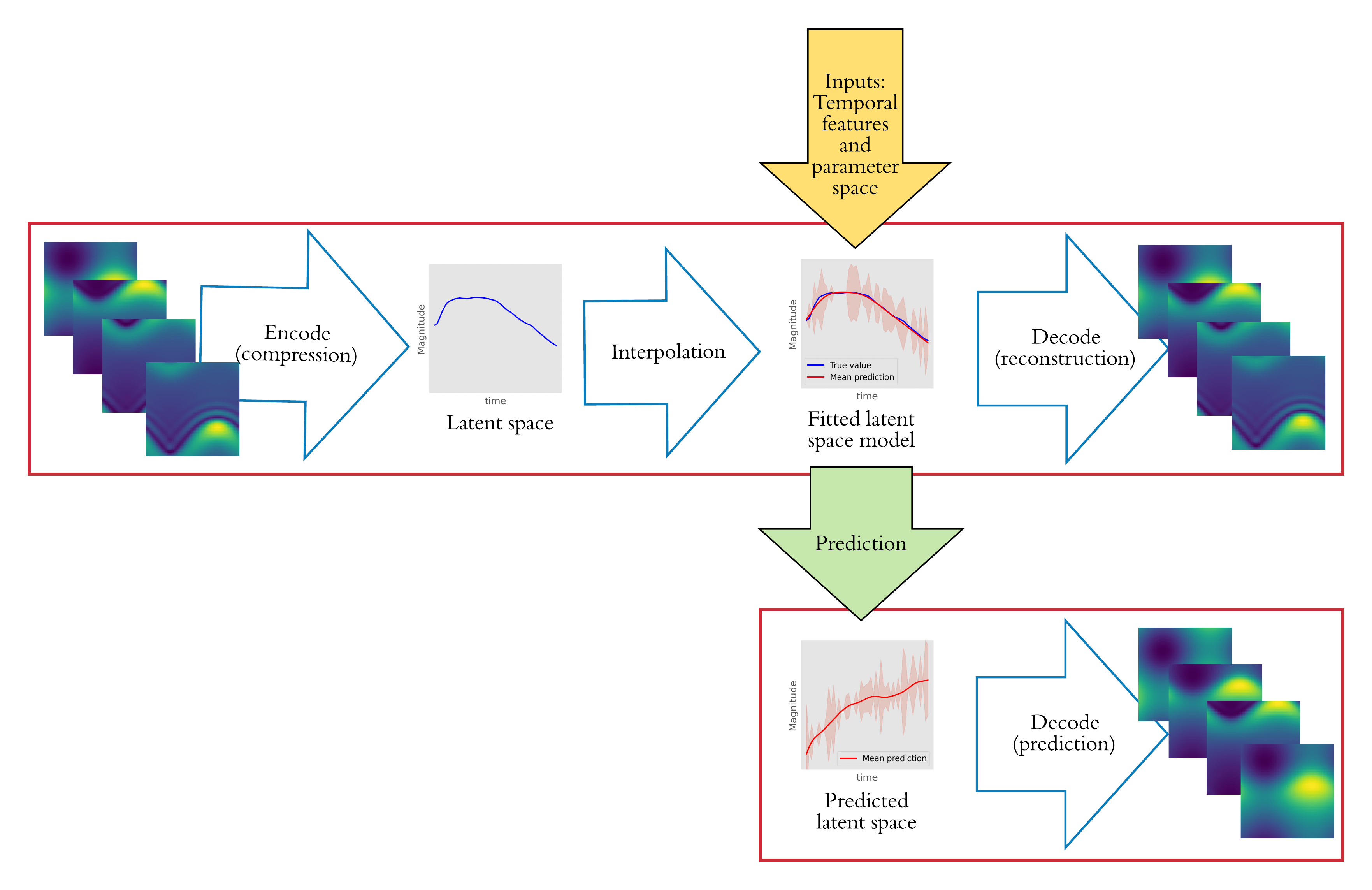}}
    \caption{Schematic of the enhanced reduced-order modeling which features 
    additional steps to Figure \ref{schematic_vanilla} involving  interpolation of the latent space and (if required) prediction for new parameters. The outputs are fed into the decoder for transformation back to the original space}
    \label{schematic_GP}
\end{figure}

The enhanced ROM methodology that includes the interpolation step is shown in Figure \ref{schematic_GP}. After we derive the latent space (in the same manner as in Section \ref{sec:compr}), we use it as input for an interpolation algorithm. The new model can be used for predictions with new sets of parameters, while the output can be further assessed and transformed back to the original space through the decoder.

There are various advantages underlying latent space interpolation. Specifically, the one that we primarily focus on in this work is the ability to interpolate and extrapolate in the parameters of interest. This is an important advantage for multiphase flow applications since new simulations can be computationally costly, and, therefore, the ability to relocate this problem into the low-dimensional latent space, where predictions are easily performed, can be valuable.
Other advantages that were explored by \cite{maulik2020latent} include interpolation in time, which can lead to finer temporal resolutions, increased interpretablility since the interpolation algorithm provides valuable visualisations of the latent space that can show the quality of the compression; and, finally, in the case where the interpolation algorithm is a Gaussian process, uncertainty quantification is also possible.

\subsubsection{Gaussian processes}\label{sec:GP}

Gaussian processes (GPs) \citep{williams2006gaussian} are generalisations of multivariate Gaussian distributions with infinite-dimensional space and a popular choice for regression \citep{williams1996gaussian} due to their versatility; the method is known as Gaussian Process regression (GPR). We mainly focus on the mean prediction of a GP that corresponds to the maximum a posteriori (MAP) estimate and use this as the decoder input.

Considering a mean function $m(\mathbf{x})$ equal to zero, a GP can be completely specified by its second-order statistics; therefore, a positive definite covariance function (otherwise known as a kernel) $k(\mathbf{x}, \mathbf{x'})$ is the only requirement.
For a GPR model, we considered a GP $f$ and noisy training observations $\mathbf{y}$ of $n$ datapoints $\mathbf{x}$ derived from the true values $f(\mathbf{x})$ with additive i.i.d.~Gaussian noise $\epsilon$ with variance $\sigma_n^2$: 
%
\begin{align}
\begin{gathered}
\mathbf{y} = f(\mathbf{x}) + \epsilon, \\
\epsilon \sim \mathcal{N}(0, \sigma_n^2),\\
f(\mathbf{x}) \sim \operatorname{GP}(0, k(\mathbf{x},\mathbf{x'})),
\end{gathered}
\end{align}
%
where $k(\cdot,\cdot)$ is the kernel. We obtain the complete GP specification by maximising the \textit{marginal likelihood}, which we can acquire by integrating the product of the Gaussian likelihood and the GP prior over $f$:
\begin{align}
\begin{gathered}
p(\mathbf{y}|\mathbf{x}) = \int_f p(\mathbf{y}|f, \mathbf{x})p(f|\mathbf{x}) \,\mathrm{d}f.
\end{gathered}
\end{align}
For testing input $\mathbf{x_\star}$ and output $\mathbf{f_\star}$, we derive the joint marginal likelihood:
\[\begin{bmatrix}
\mathbf{y} \\
\mathbf{f_\star}
\end{bmatrix}\sim \mathcal{N}\left(\begin{bmatrix}
0 \\
0
\end{bmatrix},\begin{bmatrix}
k(\mathbf{x},\mathbf{x}) + \sigma_n^2 \mathbf{I} & k(\mathbf{x},\mathbf{x_\star}) \\
k(\mathbf{x_\star},\mathbf{x}) & k(\mathbf{x_\star},\mathbf{x_\star})
\end{bmatrix}\right),
\]
where $\mathbf{I}$ is the identity matrix.
Finally, by conditioning the joint distribution on the training data and the testing inputs, we derive the predictive distribution
\begin{equation}
\mathbf{f_\star}|\mathbf{x}, \mathbf{x_\star}, \mathbf{y} \sim \mathcal{N} (\mathbf{\bar{f}}_\star, \operatorname{cov}(\mathbf{f_\star})),    
\end{equation}
where $\mathbf{\bar{f}}_\star$ and $\operatorname{cov}(\mathbf{f_\star})$ are given by
\begin{gather}
\begin{aligned}
\mathbf{\bar{f}}_\star &= k(\mathbf{x_\star},\mathbf{x})[k(\mathbf{x},\mathbf{x}) + \sigma_n^2 \mathbf{I}]^{-1}\mathbf{y} \\
\operatorname{cov}(\mathbf{f_\star}) &= k(\mathbf{x_\star},\mathbf{x_\star}) - k(\mathbf{x_\star},\mathbf{x})[k(\mathbf{x},\mathbf{x}) + \sigma_n^2 \mathbf{I}]^{-1}k(\mathbf{x},\mathbf{x_\star}).\label{predictive_dist}\\
\end{aligned}
\end{gather}

We chose a single Matérn 3/2 kernel with lengthscale $\mathbf{l}$ due to its versatility, flexibility and smoothness. Specifically, we used the automatic relevance determination (ARD) extension \cite{bishop2006pattern}, which incorporates a separate parameter for each input variable:
\begin{align}
\begin{gathered}
k(\mathbf{x}, \mathbf{x'}) = \left(1 + \frac{\sqrt{3(\mathbf{x}-\mathbf{x'})^2}}{\mathbf{l}}\right) \exp\left(-\frac{\sqrt{3(\mathbf{x}-\mathbf{x'})}}{\mathbf{l}} \right).
\end{gathered}
\label{eq:k_expression}
\end{align}
Substitution of Equation~\eqref{eq:k_expression} into Equation~\eqref{predictive_dist} yields: 
\begin{gather}
\begin{aligned}
\mathbf{\bar{f}}_\star = &\left(1 + \frac{\sqrt{3(\mathbf{x_\star}-\mathbf{x})^2}}{\mathbf{l}}\right) \exp\left(-\frac{\sqrt{3(\mathbf{x_\star}-\mathbf{x})}}{\mathbf{l}} \right)[ (1 + \sigma_n^2) \mathbf{I}]^{-1}\mathbf{y}, \\
\operatorname{cov}(\mathbf{f_\star}) = &1 - \left(1 + \frac{\sqrt{3(\mathbf{x_\star}-\mathbf{x})^2}}{\mathbf{l}}\right) \exp\left(-\frac{\sqrt{3(\mathbf{x_\star}-\mathbf{x})}}{\mathbf{l}} \right)[ (1+ \sigma_n^2) \mathbf{I}]^{-1}\\
&\left(1 + \frac{\sqrt{3(\mathbf{x}-\mathbf{x_\star})^2}}{\mathbf{l}}\right) \exp\left(-\frac{\sqrt{3(\mathbf{x}-\mathbf{x_\star})}}{\mathbf{l}} \right).\\
\end{aligned}
\end{gather}
During the reconstruction phase, we focus on the predictions that correspond to $\mathbf{\bar{f}}_\star$. 

\subsubsection{Deep Gaussian processes}\label{sec:DGP}

A Deep Gaussian Process (DGP) \citep{damianou2013deep} is a hierarchical composition of conventional GPs. In a DGP model, the data is modeled as the output of a multivariate GP, the inputs to that GP are governed by another GP and so on. In practice, DGPs are are multi-layer generalisations of GPs.
%
%
%
For the purposes of this work, we will use the doubly stochastic variational inference variant of the DGP \citep{salimbeni2017doubly}, according to which for $L$ layers we derive the joint density:
\begin{equation}\label{eq:dgp}
    p(y,\{F^l,u^l\}_{l=1}^L) =  \prod_{i=1}^{N} p(y_i|f_i^L) \prod_{l=1}^{L}p(F^l|u^l; F^{l-1},z^{l-1} )p(u^l;z^{l-1}),
\end{equation}
where $F^0 = x$. In Equation~\eqref{eq:dgp}, $x$ and $y$ are the $n$-dimensional data (inputs and outputs respectively), $F^i$ are stochastic functions with GPs as priors, $f_i^L$ is the output of the last layer, $z^i$ is a set of inducing points at layer $i$, and $u^i$ the corresponding inducing function values.
Note that in this parameterisation each of the GPs has a zero mean and the Gaussian noise is absorbed into the kernel.
By assuming that the posterior $q$ of $u^i$ is factorised between layers and $q(u^i) \sim N(m^i, S^i)$, and after marginalising the inducing variables of each layer, the marginal likelihood becomes:
\begin{equation}\label{eq:dgp_marginal}
    q(\{F^l\}_{i=1}^L) = \prod_{i=1}^L q(F^l|m^l,S^l;F^{l-1}, Z^{l-1}) = \prod_{i=1}^L N(F^l|\bar{\mu}^l,\bar{\Sigma}^l).
\end{equation}
For new predictions the following equation applies:
\begin{equation*}
    q(f_*^L) = \frac{1}{V}\sum_{v=1}^V q(f_*^L|m^L,S^L;f_*^{(v)(L-1)}, Z^{L-1}),
\end{equation*}
where $f_*^{(v)(L-1)}$ are $V$ samples from Equation~\eqref{eq:dgp_marginal}.
For the implementation of GPs and DGPs we used the GPyTorch library \citep{gardner2018gpytorch}.

\subsubsection{Long short-term memory networks}

The Long short-term memory networks (LSTMs) \citep{hochreiter1997long} are a special case of recurrent neural networks (RNNs), a class of neural networks that account for sequential data, thus being particularly useful for problems with temporal components. LSTMs, specifically, use gated cells that allow information transfer from both the recent past (short-term memory) and the distant past (long-term memory). This is an advantage of LSTMs compared to other RNN variations in terms of the results quality and also presents a solution to practical problems such as vanishing gradients (very small gradients during backpropagation that can render neurons inactive).
A typical LSTM cell consists of:
\begin{align*}
    \text{forget gate} &: f_t = \sigma_g(W_fx_t + U_fh_{t-1} + b_f) \\
    \text{input gate} &: i_t = \sigma_g(W_ix_t + U_ih_{t-1} + b_i)  \\
    \text{output gate} &: o_t = \sigma_g(W_ox_t + U_oh_{t-1} + b_o) \\
    \text{cell input} &: \bar{c_t} = \sigma_c(W_cx_t + U_ch_{t-1} + b_c) \\
    \text{cell state} &: c_t = f_t \odot c_{t-1}  + i_t \odot  \bar{c_t} \\
    \text{hidden state} &: h_t = o_t \odot c_t,
\end{align*}
where $t$ denotes the time, $x_t$ is the input, $W_*$, $U_*$ and $b_*$ are the weights of the input and recurrent connection matrices and bias terms of the quantity $*$ respectively, $\odot$ is the element-wise product,  $\sigma_g$ is the sigmoid function and $\sigma_c$ is the hyperbolic tangent function.

\section{Applications}\label{sec:Apps}
We apply the ROM analysis pipeline to three fluid-dynamic simulation applications with increasing complexity: (i) an advection–diffusion equation, (ii) a falling film flow, and (iii) multi-component polymer precipitation governed by a Cahn–Hilliard equation. We demonstrate how the methodology variants perform on simulation data-sets via  appropriate visualisations and metrics.

\subsection{Advection–diffusion}\label{sec:adv}
Advection–diffusion equations expressed as
\begin{equation}\label{eq:Burgers2D}
    \frac{\partial c}{\partial t} + \overrightarrow{v} \cdot \overrightarrow{\nabla} c = D \nabla^2 c
\end{equation}
describe physical phenomena where mass, momentum, and energy are transported within a physical domain advectively by bulk motion and diffusively in response to differences in  chemical potential and/or temperature. 
%
%
%
Here, $c$ is a concentration, $D$ is a constant diffusion coefficient, $\overrightarrow{v} = (v_x(x,y), v_y(x,y))$ is the velocity field, and $\overrightarrow{\nabla}$ denotes the gradient operator. Note that for a constant velocity, $U_c$, and an appropriate characteristic length scale, $L_c$, 
one can scale space and time on $L_c$ and $L_c/U_c$, respectively, to arrive at
\begin{equation}
\frac{\partial c}{\partial t}+
\frac{\partial c}{\partial x}+
\frac{\partial c}{\partial y}=
\frac{1}{Pe}
\left(\frac{\partial^2 c}{\partial x^2} + \frac{\partial^2 c}{\partial y^2}
\right),
\label{eq:advection_diffusion}
\end{equation}
where $Pe=U_c L_c/D$ is a Péclet number.

To generate a characteristic advection–diffusion simulation data-set, we implement and solve Equation~\eqref{eq:advection_diffusion} using the framework of \cite{bar2019learning} and \cite{zhuang2020learned}.
For training purposes, we generate \textcolor{black}{$20$} simulations for different values of $Pe^{-1}$ ranging between \textcolor{black}{$0.05$} and \textcolor{black}{$0.15$}, representing a blob of inert tracer placed in a constant-velocity fluid field. For each simulation, $50$ evenly spaced time-snapshots with $64\times 64$ resolution are obtained to construct the full data-set.
We split the simulations into \textcolor{black}{$19$} cases for training and 
1 for testing, focusing on the \textit{extrapolation} problem. 
\begin{table}[b]
\caption{Evaluation metrics for the POD, CAE and VAE models for the advection–diffusion problem. The best performance for each metric (i.e. MAE, MSE) is highlighted in bold.}
\centering
\begin{tabular}{l ccc}
    \toprule
             & \multicolumn{3}{c}{POD} \\
    \cmidrule(lr){2-4}
    Metric/Model   & GP                   & DGP           & LSTM                  \\ \midrule
    MAE  & \num{2.51355e-5}  & \num{4.32963e-5}   & \num{0.00203} \\
    MSE  & \num{0.00264}    & \num{0.00292}  & \num{0.01594}  \\ \bottomrule
    \toprule
             & \multicolumn{3}{c}{CAE} \\
    \cmidrule(lr){2-4}
    Metric/Model   & GP                   & DGP           & LSTM                  \\ \midrule
    MAE  & \num{7.20011e-7}  & \num{1.28770e-6}   & \textcolor{blue}{\textbf{\num{4.83279e-7}}} \\
    MSE  & \num{4.48591e-5}    & \num{6.12765e-5}  & \num{4.98987e-5}  \\ \bottomrule
    
    \toprule
             & \multicolumn{3}{c}{VAE} \\
    \cmidrule(lr){2-4}
    Metric/Model   & GP                   & DGP           & LSTM                  \\ \midrule
    MAE  & \num{6.31853e-7}  & \num{1.22549e-6}   & \num{4.87542e-7} \\
    MSE  & \textcolor{blue}{\textbf{\num{4.27802e-5}}}    & \num{6.30656e-5}  & \num{4.57708e-5}  \\ \bottomrule
\end{tabular}
\label{table:adv_metrics_all}
\end{table}
%
The data are fed into the three algorithms presented in Section \ref{sec:compr}. We train the different compression algorithms for \textcolor{black}{$4$} degrees of freedom (DOF). For an extensive analysis on how different DOF affect the result, we refer the reader to the experiment section in \cite{maulik2020latent}.
For the encoders of the CAE and VAE we use $5$ convolutional and pooling layers and an equal number of unpooling and upsampling layers for the decoders. We also train the data in mini-batches of $4$, and we use $1,000$ epochs and early stopping based on a validation set that consists of \textcolor{black}{$10\%$} of the training data. We use TensorFlow \citep{abadi2016tensorflow} for autoencoder implementations.
For the GPs we use an ARD Matérn 3/2 kernel with two-dimensional lengthscale (one dimension corresponds to the diffusion coefficient and the other to the temporal parameter).
For the DGPs we use \textcolor{black}{$128$} inducing points, \textcolor{black}{$2$} layers, each of which comprises of \textcolor{black}{a single} GP with a Matérn 3/2 kernel.
Finally, for the LSTM architecture, we use \textcolor{black}{$3$} cells with \textcolor{black}{$50$} neurons in each cell, mini-batches of size \textcolor{black}{$32$}, early stopping based on \textcolor{black}{$10\%$} of the training data that are set aside, and a time window of \textcolor{black}{$10$} points for the forecasts, following a similar approach to \cite{maulik2021reduced}.
All the interpolation algorithms use the $Pe$ value as an additional input to perform the latent space interpolation.

\begin{figure*}[t]
    \centering
    \includegraphics[width=\textwidth]{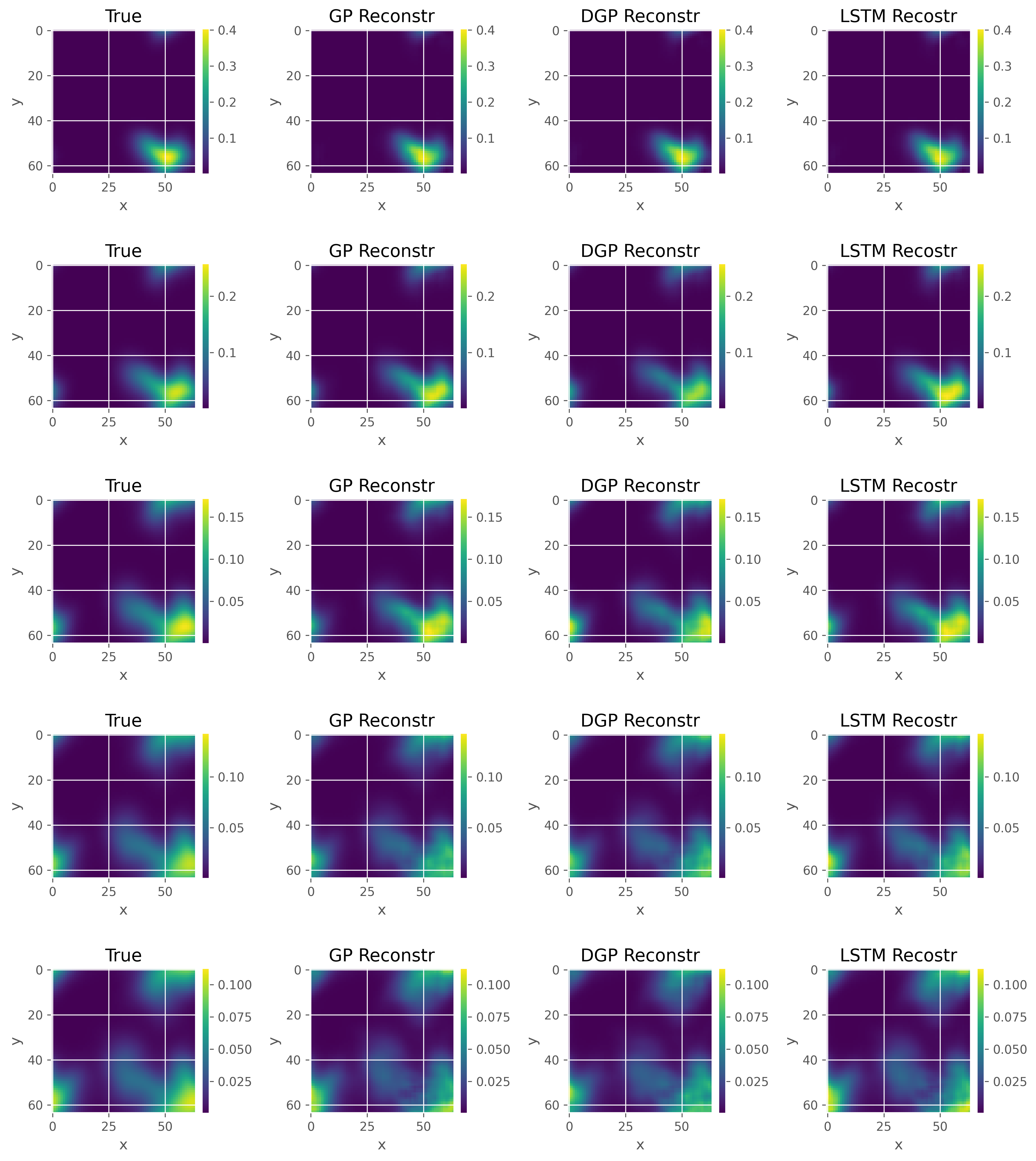}
    \caption{True and reconstructed frames for the VAE-related methods applied to the advection–diffusion problem. The rows correspond to time-steps $11, 21, 31, 41$ and $50$, respectively}
    \label{fig:adv_VAE}
\end{figure*}

\begin{figure*}
    \centering
    \includegraphics[width=0.75\textwidth]{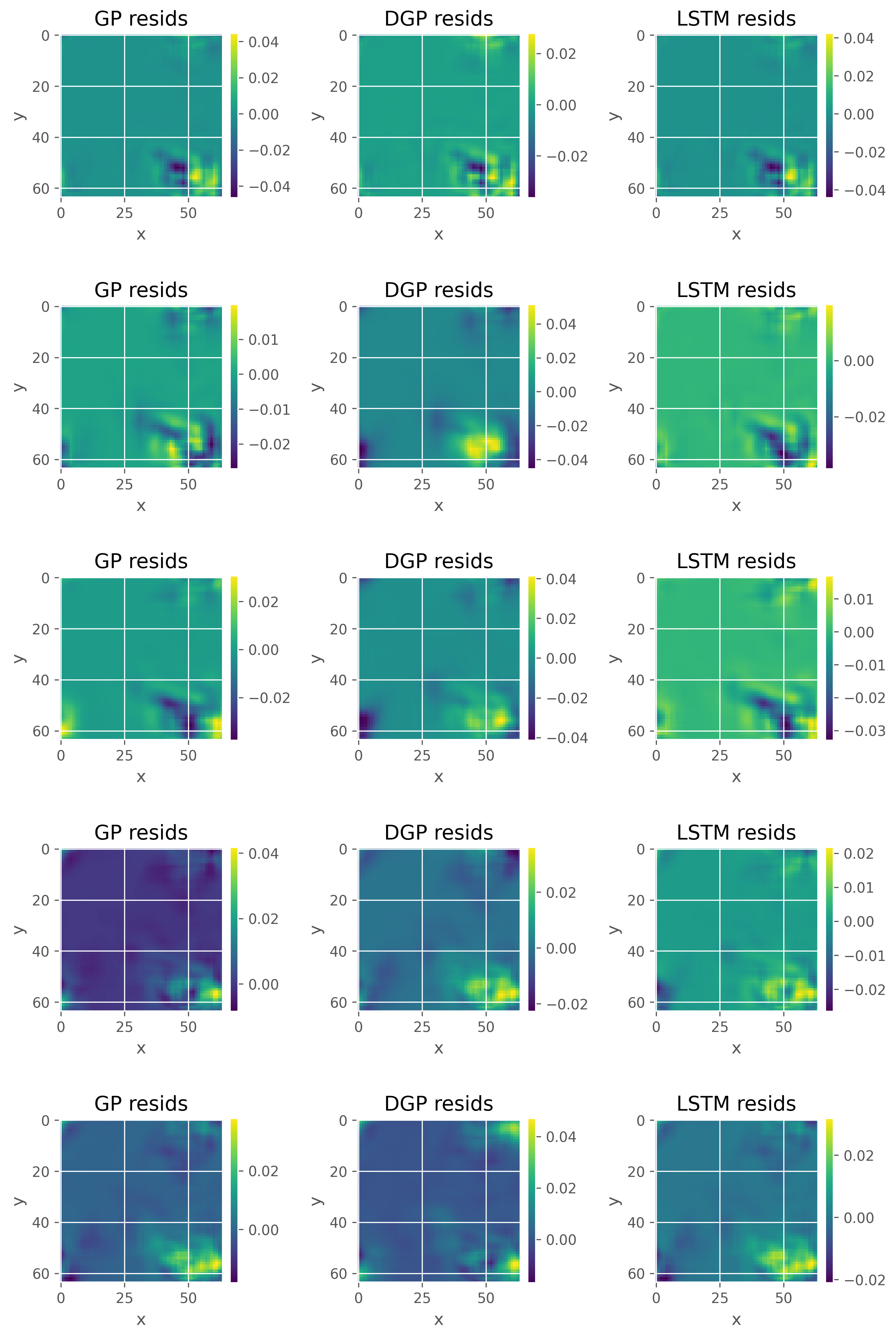}
    \caption{Residual plots of the VAE-related methods applied to the advection–diffusion problem. The rows correspond to time-steps $11, 21, 31, 41$ and $50$, respectively}
    \label{fig:adv_VAE_Resids}
\end{figure*}

To assess the different methods, we use two metrics: the mean squared error (MSE) and the mean absolute error (MAE). We remind the reader that output frames derived directly from the compression algorithms (in a typical ROM assessment fashion) are not possible to derive, since we make predictions for $Pe$ 
values outside the training set.
%
It is clear from Table \ref{table:adv_metrics_all} that POD and DGP are consistently the worst performing of the compression and interpolation algorithms, respectively (with a few exceptions, such as DGP outperforming LSTM for the POD case); 
the conclusion associated with POD was one also reached by \cite{maulik2020latent}. As for the DGP conclusion, it appears that the additional complexity of using the DGP is unwarranted when applied to relatively small and simple data-sets, and that the corresponding model may become over-parameterised. Instead, the standard GP is flexible enough to capture the variability of the data obviating the need for a DGP.
It is also unclear based on the results presented which of the two autoencoders and remaining interpolation algorithms performs better. The MAE favours all the LSTM variations, and specifically the CAE–LSTM combination, while the MSE is lower for the GP variations and especially VAE–GP, indicating that the LSTM offers lower average error and the GP fewer error spikes. The error values, however, are very small signifying the overall effectiveness of the methodology.

In Figures \ref{fig:adv_VAE} and \ref{fig:adv_VAE_Resids} we show indicative input frames (from top to bottom: $11$ $21$, $31$, $41$, $50$), along with the different VAE-based reconstructions and the corresponding residual plots, respectively. Specifically, the first column in Figure \ref{fig:adv_VAE} corresponds to the true frames, the second to the GP reconstruction, the third to the DGP reconstruction, and the final to the LSTM reconstruction. The scaling is common for each row. Figure \ref{fig:adv_VAE_Resids} uses the last three cases for the corresponding residuals. In order to show the differences among these plots we use flexible scaling.
In terms of the actual frames we can see that the reconstruction is successful and it is hard to detect differences from the originals, even in the case of the DGP, which is the interpolation algorithm that performed more poorly, with some minor exceptions in the bottom frames.
Regarding the residuals, we can see the similar error patterns generated, though, the scale makes it clear that the average error is larger for some of the frames in the GP and most in the DGP cases, when compared to the LSTM column.


\begin{figure}
\centering
\begin{subfigure}{.3\textwidth}
  \centering
  \includegraphics[width=4cm,height=3cm]{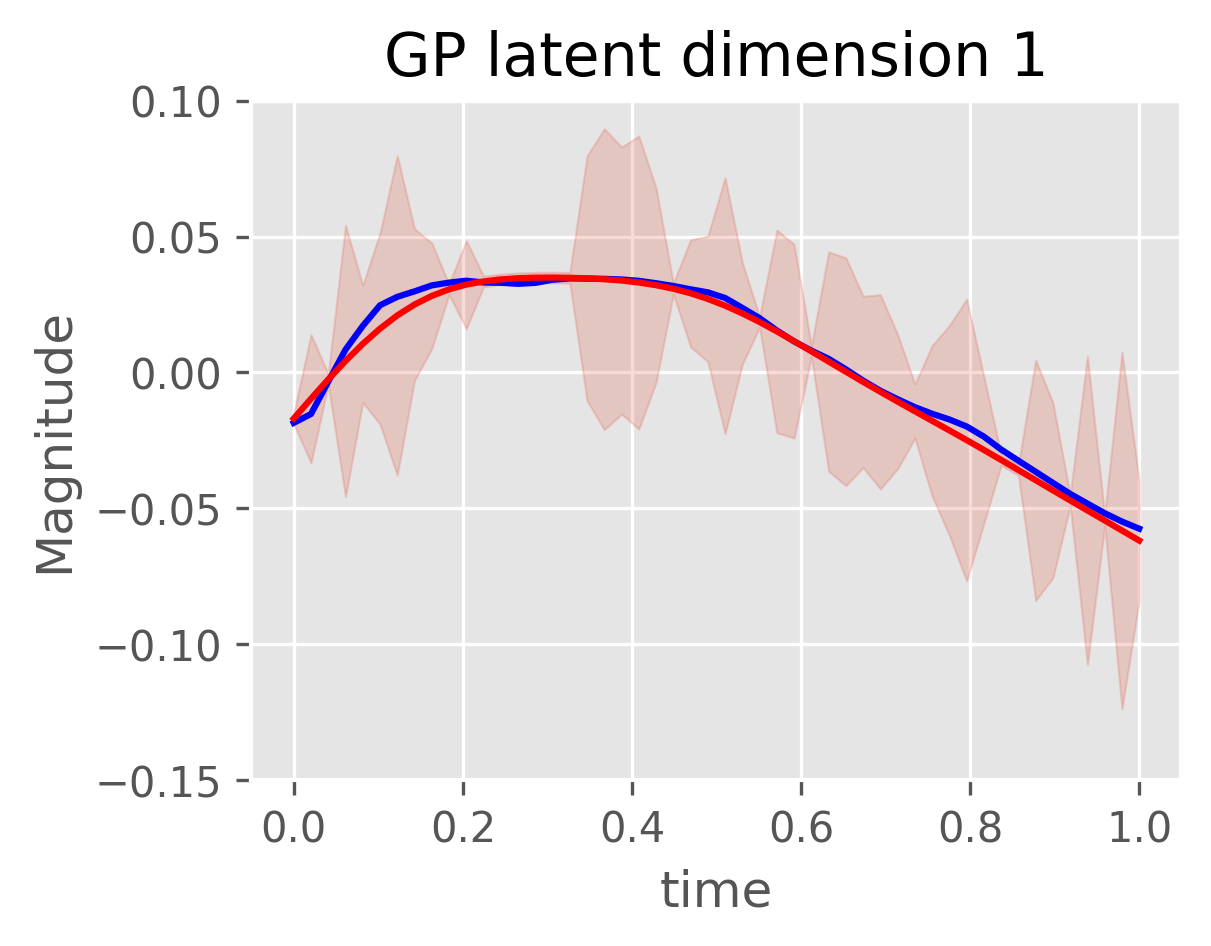}
  \caption{}
  \label{fig:sub11}
\end{subfigure}%
\begin{subfigure}{.3\textwidth}
  \centering
  \includegraphics[width=4cm,height=3cm]{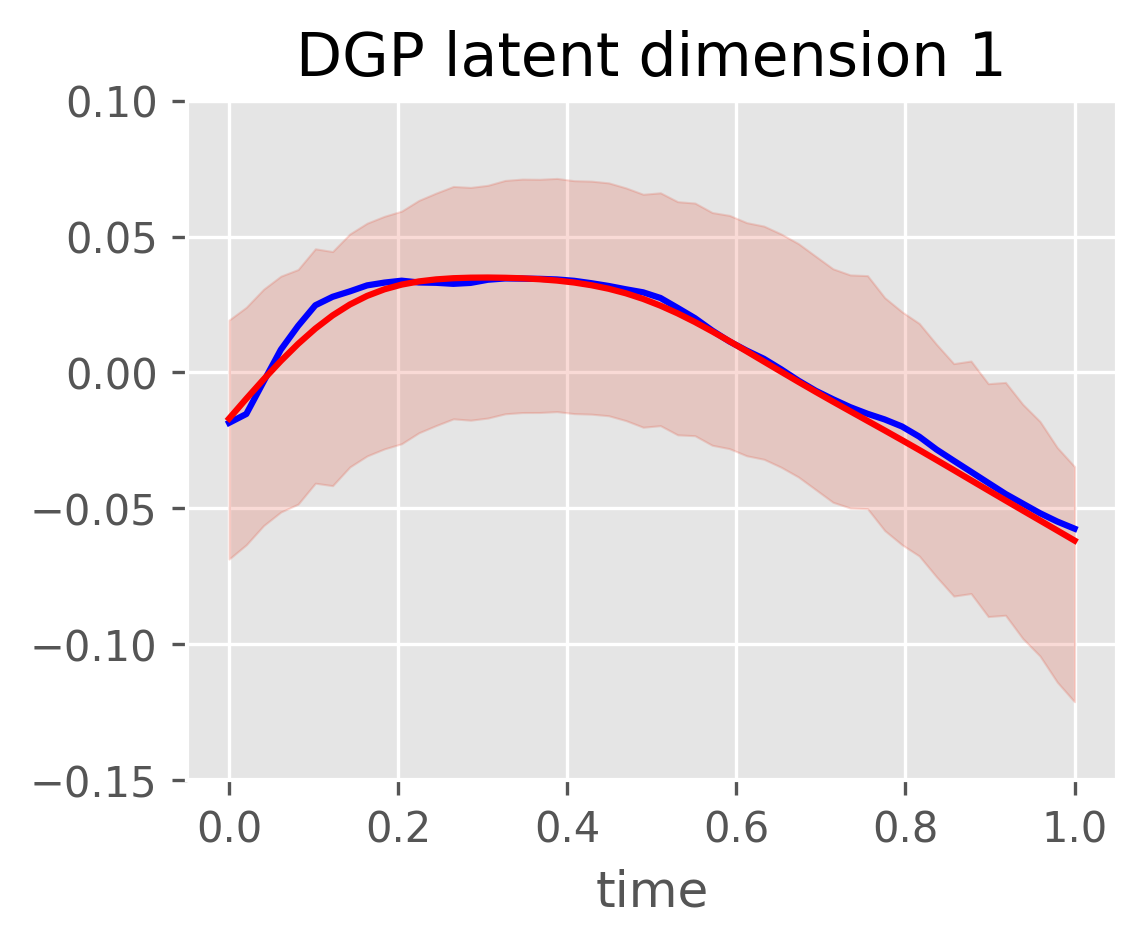}
  \caption{}
  \label{fig:sub22}
\end{subfigure}%
\begin{subfigure}{.3\textwidth}
  \centering
  \includegraphics[width=4cm,height=3cm]{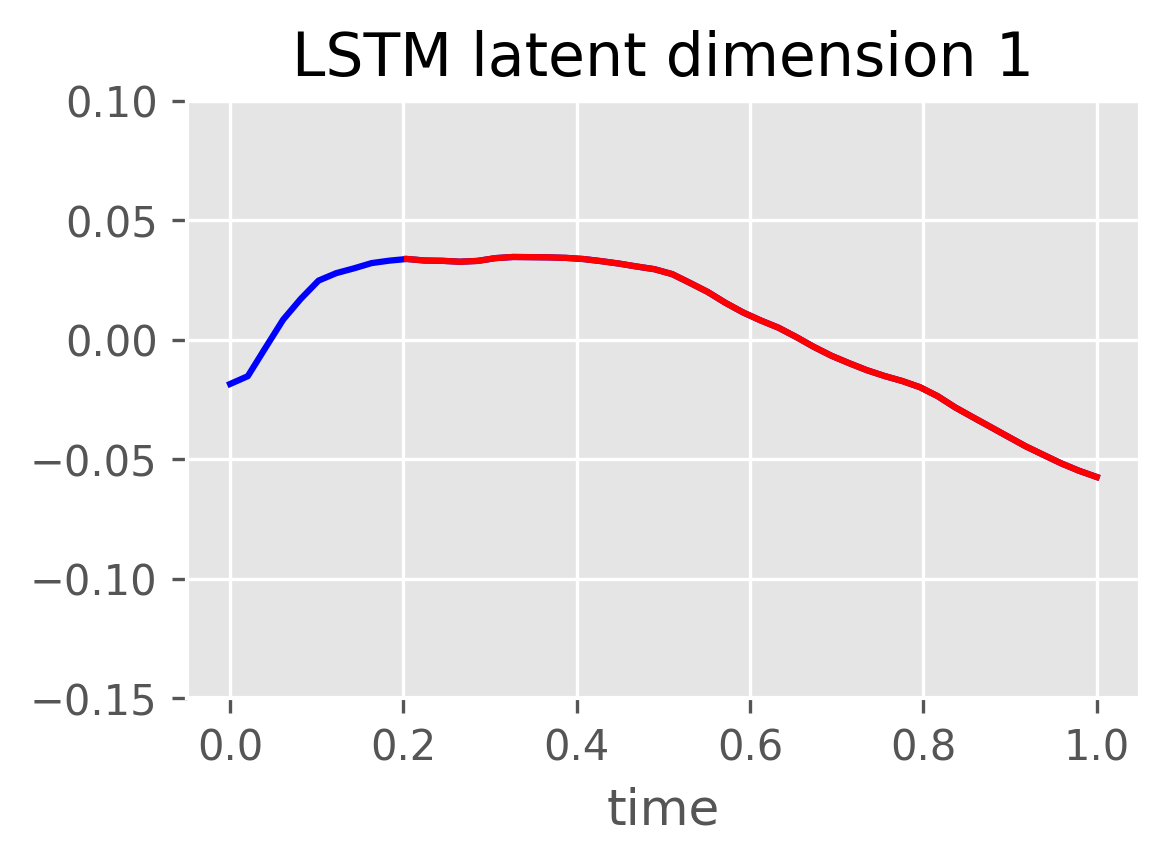}
  \caption{}
  \label{fig:sub33}
\end{subfigure}
\caption{Comparison of the interpolation techniques in the latent space for advection–diffusion}
\label{fig:adv_latent}
\end{figure}

    

\begin{table}[b]
\caption{Computational cost (in seconds) for all compression and interpolation algorithms.}
\centering
\begin{tabular}{ccc}
    \toprule
    \multicolumn{3}{c}{POD} \\
    \midrule
    GP                   & DGP           & LSTM\\
    \num{9.1e1}  & \num{5.96e2}   & \num{4.2e1} \\
    \bottomrule
    
    \toprule
    \multicolumn{3}{c}{CAE} \\
    \midrule
    GP                   & DGP           & LSTM\\
    \num{1.46e2}  & \num{6.19e2}   & \num{3.9e1} \\
    \bottomrule
 
    \toprule
    \multicolumn{3}{c}{VAE} \\
    \midrule
    GP                   & DGP           & LSTM\\
    \num{1.14e2}  & \num{4.3e2}   & \num{5.6e1} \\
 \bottomrule

\end{tabular}
\label{table:adv_time}
\end{table}

In Figure \ref{fig:adv_latent}, we gain insights into the practical differences between the three interpolation algorithms by showing the corresponding representations of the first dimension of the latent space. The blue line shows the actual first dimension from the VAE latent space, the red line the mean for the GP-based algorithms for the first two and the output of the LSTM for the third, and the light red shading the confidence intervals around $2$ standard deviations from the mean.
Starting from the GP (Figure \ref{fig:sub11}) we can see a very good fit and an inconsistent uncertainty level. Generally, the uncertainty is narrower close to the times where data are available and wider everywhere else. The DGP case (Figure \ref{fig:sub22}) shows a very similar mean result, but the over-parameterisation resulted to a wide uniform credible interval along the time axis. Finally, the LSTM (Figure \ref{fig:sub33}) shows an almost perfect fit to the latent space. The missing line before $t=0.2$ of the plot corresponds to the timewindow of $10$ points that is required for the forecast.

The computational cost for all the algorithms examined in this section is shown in Table \ref{table:adv_time}. The compression algorithm is run first and its cost is common for all relative pipelines regardless of the choice of the interpolation. POD is almost instantaneous, while the autoencoders require a similar amount of time with each other. The time to solution for the interpolation algorithms differ based on the compression algorithm they follow, but the scale remains approximately the same. Specifically, LSTMs are consistently the fastest, followed by GPs, while DGPs are the slowest. It is worth noting that different parameterisations of the algorithms can potentially lead to significant differences with respect to computational costs, within each class of compression and interpolation.



\begin{table}[b]
\caption{Evaluation metrics for the POD, CAE and VAE models. The best performance for each metric (i.e. MAE, MSE) is highlighted in bold.}
\centering
\begin{tabular}{l ccc}
    \toprule
             & \multicolumn{3}{c}{POD} \\
    \cmidrule(lr){2-4}
    Metric/Model   & GP                   & DGP           & LSTM                  \\ \midrule
    MAE  & \num{0.04865}  & \num{0.04819}   & \num{0.04055} \\
    MSE  & \num{0.17284}    & \num{0.15188}  & \num{0.16617}  \\ \bottomrule
    \toprule
             & \multicolumn{3}{c}{CAE} \\
    \cmidrule(lr){2-4}
    Metric/Model   & GP                   & DGP           & LSTM                  \\ \midrule
    MAE  & \num{5.45812e-5}  & \num{3.28181e-4}   & \textcolor{blue}{\textbf{\num{1.90501e-5}}} \\
    MSE  & \num{0.00056}    & \num{0.00149}  & \textcolor{blue}{\textbf{\num{0.00036}}}  \\ \bottomrule
    
    \toprule
             & \multicolumn{3}{c}{VAE} \\
    \cmidrule(lr){2-4}
    Metric/Model   & GP                   & DGP           & LSTM                  \\ \midrule
    MAE  & \num{0.00013}  & \num{0.00069}   & \num{5.46090e-5} \\
    MSE  & \num{0.00079}    & \num{0.00229}  & \num{0.00079}  \\ \bottomrule
\end{tabular}
\label{table:films_metrics_all}
\end{table}

\subsection{Falling films}
We focus in this application on two-dimensional falling film flows, of importance to engineering applications including separation and heat removal units \citep{rohlfs2018wavemaker}. We apply the modelling framework of \cite{scheid2006wave}, which uses an averaged version of the Navier–Stokes equations together with Pade approximants; this yields a three-field equation system for the film thickness, $h$, and flow rates in the streamwise direction $x$, $q^{x}$, and spanwise one $y$, $q^{y}$, expressed by
\begin{equation}
    \frac{\partial h}{\partial t}=-\frac{\partial q^{x}}{\partial x}-\frac{\partial q^{y}}{\partial y},
    \label{eq:ffh}
\end{equation}
\begin{equation}
    \delta \frac{\partial q^{x}}{\partial t}=\delta \frac{\partial{{\cal{I}}^{2D}}}{\partial x}+{\cal{G}}\left(\frac{5}{6}h-\frac{5}{2h^2}\frac{\partial q^x}{\partial x}+\delta \frac{\partial {\cal I}^{3D}}{\partial x}
    +\eta \left[\frac{\partial {\cal D}^{2D}}{\partial x}+\frac{\partial{{\cal{D}}^{3D}}}{\partial x}\right]+\frac{5h}{6}\frac{\partial {\cal{P}}}{\partial  x}
    \right),
    \label{eq:ffqx}
\end{equation}
\begin{equation}
    \delta \frac{\partial q^{y}}{\partial t}=\delta \frac{\partial{{\cal{I}}^{2D}}}{\partial y}-\frac{5}{2h^2}\frac{\partial q^y}{\partial y}+\delta \frac{\partial {\cal I}^{3D}}{\partial y}
    +\eta \left[\frac{\partial {\cal D}^{2D}}{\partial y}+\frac{\partial{{\cal{D}}^{3D}}}{\partial y}\right]
    +\frac{5h}{6}\frac{\partial {\cal{P}}}{\partial  y},
    \label{eq:ffqy}
\end{equation}
where ${\cal{G}}=\left[1-(\delta/70)h^3\frac{\partial h}{\partial x}\right]^{-1}$, ${\cal {I}}$ and ${\cal{D}}$ represent a collection of terms related to inertia and viscous dissipation, defined in equations (B1b) and (B1c) in \cite{scheid2006wave}, and ${\cal{P}}=-\zeta h +h\left(\frac{\partial^2 h}{\partial x^2}+\frac{\partial^2h}{\partial y^2}\right)$ represents contributions to the pressure due to gravity and capillarity. The parameters $\delta$, $\eta$, and $\zeta$ respectively represent the relative significance of inertia, gravity, and capillarity viz. equations (2.9) and (2.10) in \cite{scheid2006wave}; the parameter $\delta$ is referred to as a reduced Reynolds number.

\begin{figure*}
    \centering
    \includegraphics[width=\textwidth]{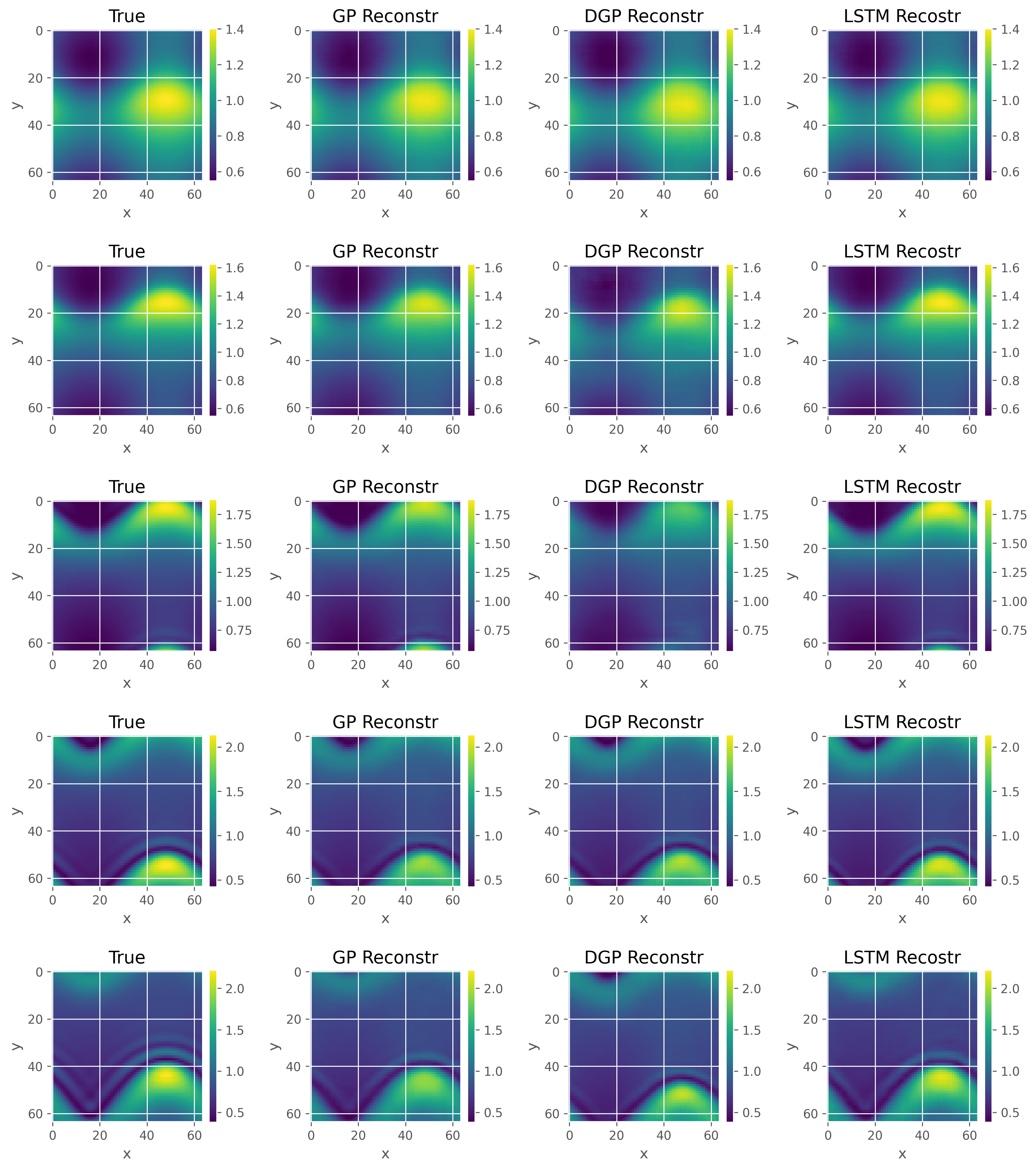}
    \caption{True and reconstructed frames for the CAE-related methods employed in the falling film problem. The rows correspond to time-steps $11, 21, 31, 41$ and $50$ respectively}
    \label{fig:films_CAE}
\end{figure*}

\begin{figure*}
    \centering
    \includegraphics[width=0.75\textwidth]{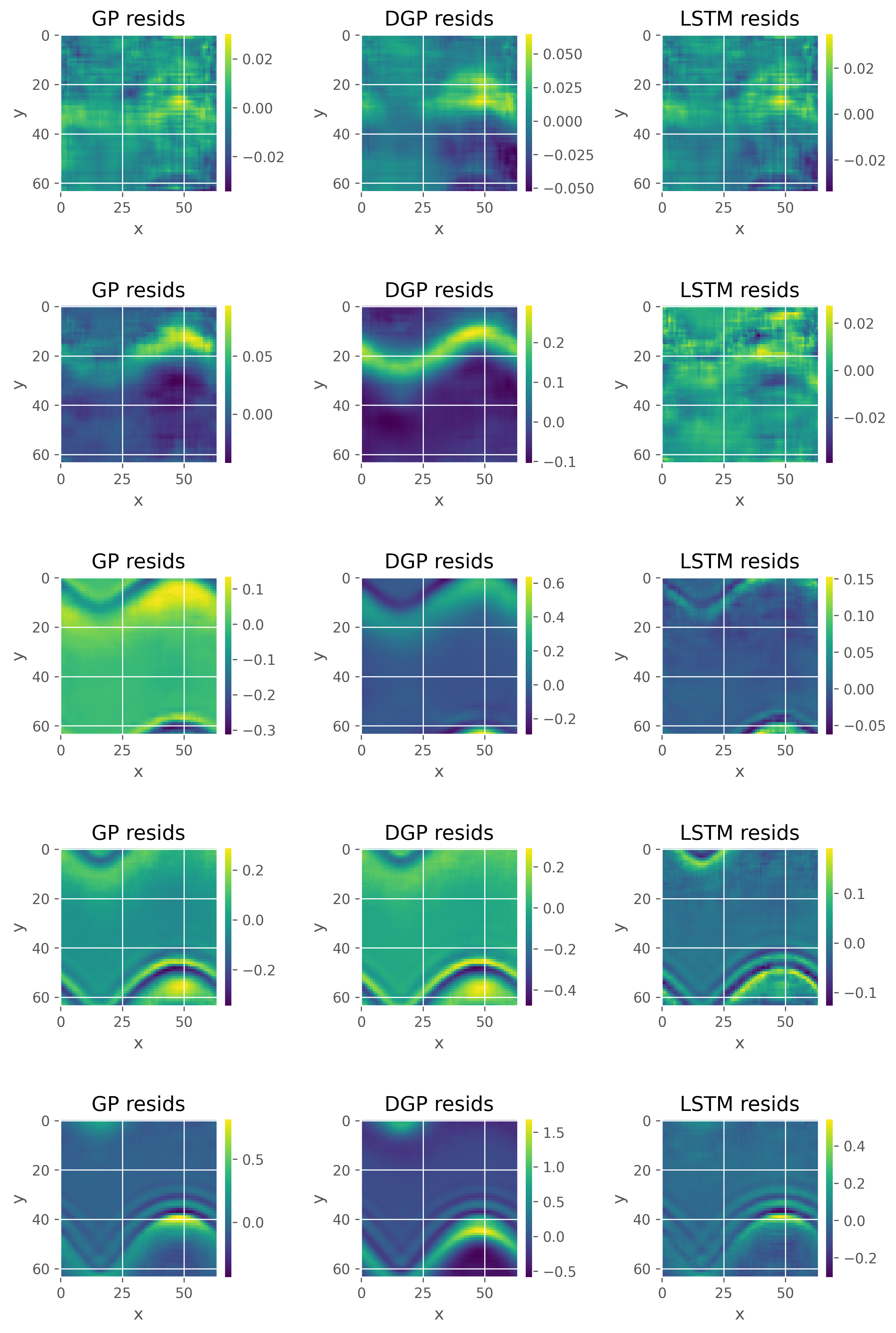}
    \caption{Residual plots of the CAE-related methods used in the falling film problem. The rows correspond to time-steps $11, 21, 31, 41$ and $50$ respectively}
    \label{fig:films_CAE_Resids}
\end{figure*}

Data were obtained from numerical solutions of Equations (\ref{eq:ffh}) to (\ref{eq:ffqy}) simulated using a custom port of WaveMaker \citep{rohlfs2018wavemaker} implemented in the Julia language \citep{bezanson2017julia}. 
%
%
The general structure for this application has a significant degree of similarity with the one applied in Section \ref{sec:adv}, but with a few key differences: first, the simulations are generated based on the reduced Reynolds number $\delta$ that takes $20$ equi-spaced values in the range of $1$ to $69$. This is also the parameter used for the interpolation algorithms. Second, in this application we focus on the \textit{interpolation} problem. Specifically, we use the $6$th simulation (that corresponds to $\delta = 25.5$) for testing and the rest for training. Finally, we use $5$ DOF.


The metrics for all the results are shown in Table \ref{table:films_metrics_all}. Similar to the first advection–diffusion application, the POD and the DGP underperform when compared to the other methods. For this data-set the combination of CAE and LSTM was the best performing based on both the MAE and the MSE metrics.
As with the advection–diffusion application, finding differences between the frames for the interpolation algorithms paired with the CAE compression in Figure \ref{fig:films_CAE} is difficult, regardless of the significant differences shown by the metrics. The only notable exception is the DGP reconstruction that shows slight deviations from the true concentration profile. It is interesting to note that for all methods in Figure \ref{fig:films_CAE_Resids}, and particularly the LSTM, the error structure in the upper frames (early times) seems random, whereas in the bottom frames (late times) we can clearly see the general structure of the concentration profile. This indicates that it is easier for the methodology to reconstruct the frames from the beginning of the simulations (where all training simulations are similar) rather than the ending (where deviations among simulations are visibly different from each other due to the effect of the different reduced Reynolds numbers).

\begin{table}[b]
\caption{Evaluation metrics for the POD, CAE and VAE models. The best performance for each metric (i.e. MAE and MSE) is highlighted in bold.}
\centering
\begin{tabular}{l ccc}
    \toprule
             & \multicolumn{3}{c}{POD} \\
    \cmidrule(lr){2-4}
    Metric/Model   & GP                   & DGP           & LSTM                  \\ \midrule
    MAE  & \num{0.00756}  & \num{0.00808}   & \num{0.03973} \\
    MSE  & \num{0.06581}    & \num{0.06553}  & \num{0.15826}  \\ \bottomrule
    \toprule
             & \multicolumn{3}{c}{CAE} \\
    \cmidrule(lr){2-4}
    Metric/Model   & GP                   & DGP           & LSTM                  \\ \midrule
    MAE  & \textcolor{blue}{\textbf{\num{7.50197e-5}}}  & \num{0.02313}   & \num{0.00082} \\
    MSE  & \textcolor{blue}{\textbf{\num{0.00084}}}   & \num{0.01698}  & \num{0.00277}  \\ \bottomrule
    
    \toprule
             & \multicolumn{3}{c}{VAE} \\
    \cmidrule(lr){2-4}
    Metric/Model   & GP                   & DGP           & LSTM                  \\ \midrule
    MAE  & \num{0.00081}  & \num{0.02814}   & \num{0.00111} \\
    MSE  & \num{0.00317}    & \num{0.01871}  & \num{0.00414}  \\ \bottomrule
\end{tabular}
\label{table:precip_metrics_all}
\end{table}

\begin{figure*}
    \centering
    \includegraphics[width=\textwidth]{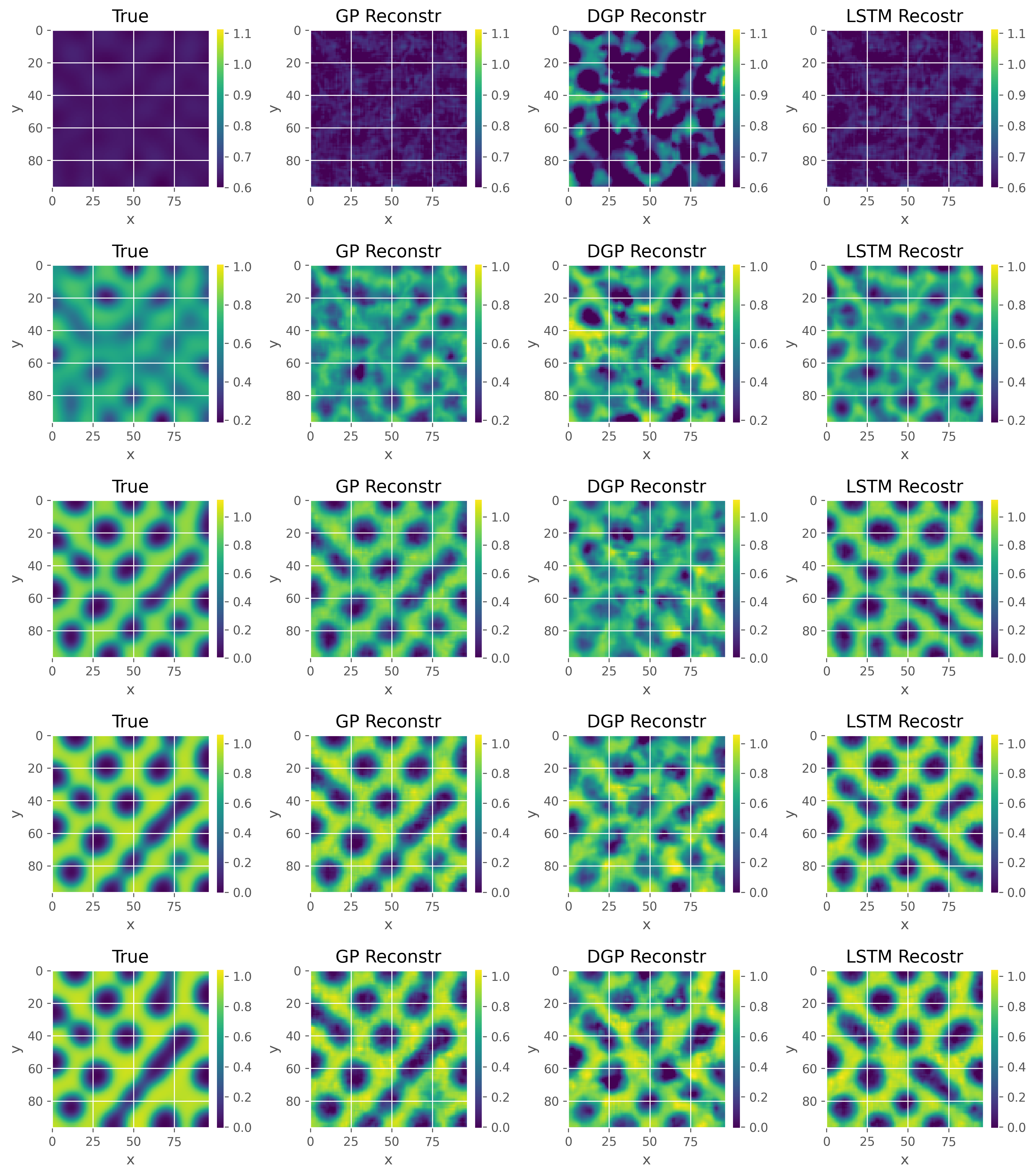}
    \caption{True and reconstructed frames for the CAE-related methods used in the polymer precipitation problems. The rows correspond to time-steps $11, 21, 31, 41$ and $50$ respectively}
    \label{fig:precip_CAE}
\end{figure*}

\begin{figure*}
    \centering
    \includegraphics[width=0.75\textwidth]{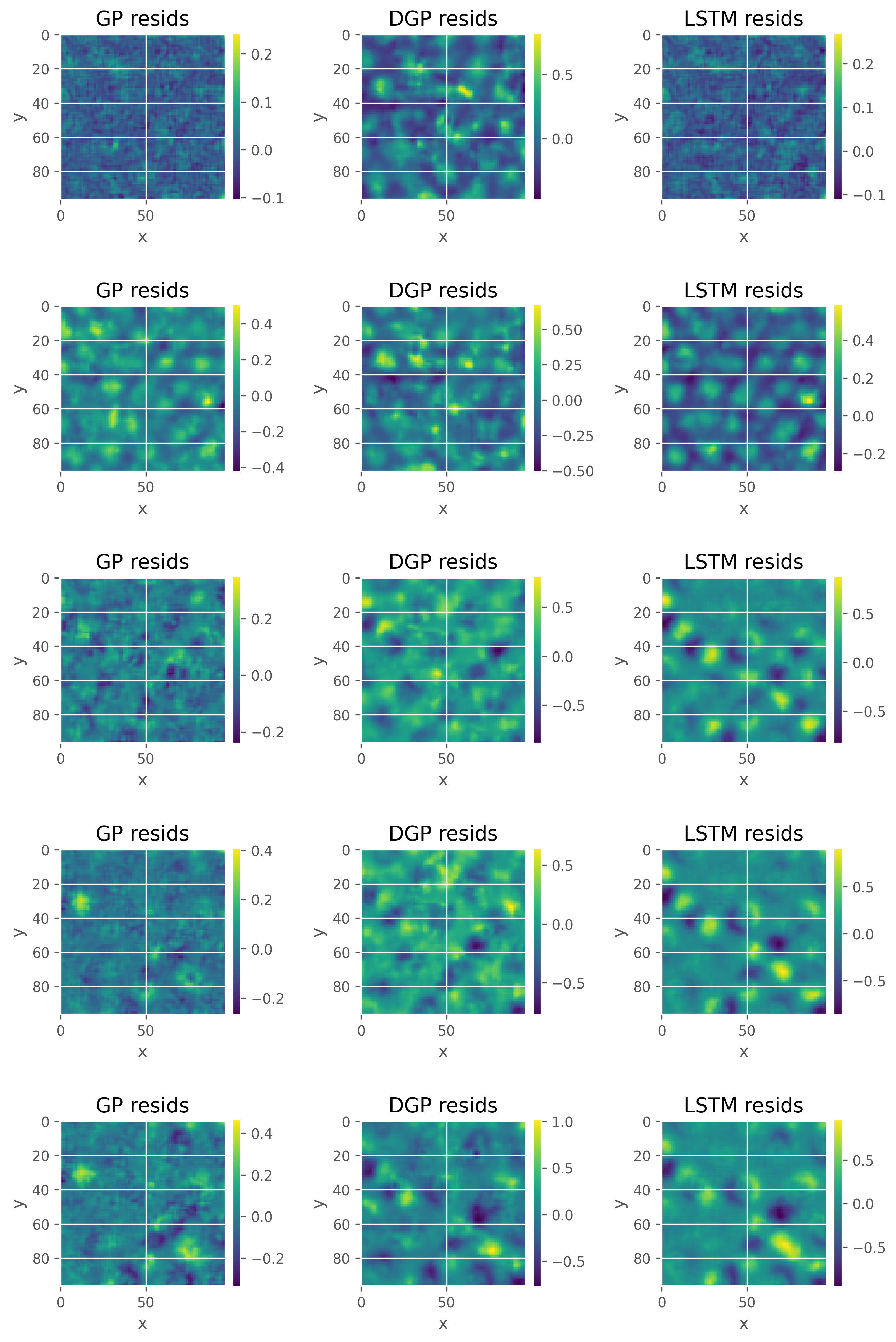}
    \caption{Residual plots of the CAE-related methods in the polymer precipitation problems. The rows correspond to time-steps $11, 21, 31, 41$ and $50$ respectively}
    \label{fig:precip_CAE_Resids}
\end{figure*}

\subsection{Multicomponent polymer precipitation}\label{sec:precip}
We now apply the analysis pipeline to polymer precipitation dynamics, of importance to engineering design problems in high-performance plastics and membrane systems. The complex dynamics and rich pattern formation were considered recently by \cite{inguva2020numerical} and \cite{inguva2021softmatter} (see also references therein) who used Cahn–Hilliard theory to model and simulate the spatio–temporal evolution of the emergent phase separation patterns; the relevant equations for a binary polymer blend are expressed by
\begin{equation}
\frac{\partial \phi}{\partial t} - \nabla \cdot \left(M \nabla\mu\right)  = 0,
\end{equation}
where $\phi$  represents the volume fraction of one of the polymers in the blend, $M$ is a constant mobility parameter, and $\mu$ is a generalised chemical potential, which can be derived from the variational derivative of the Gibbs free energy functional: 
\begin{equation}
    \mu =  \frac{d f}{d \phi} - \lambda \nabla^{2}\phi;
\end{equation}
here, $f$ denotes the homogeneous contribution to the Gibbs free energy per monomer, which is a non-convex function of 
$\phi$, and $\lambda$ is a gradient free energy parameter. 
Numerical solutions of the above equations are obtained subject to Neumann conditions: 
%
%
%
%
\begin{align*}
\nabla \mu \cdot {\mathbf{n}} & =0, \\
\nabla c \cdot {\mathbf{n}} & =0,
\end{align*}
in which $\mathbf{n}$ is the outward-directed boundary normal. 
%
%



For training purposes, we generate $20$ simulations based on values of $\lambda$ within the range $0.01$ to $0.0575$, which is also the parameter used for the interpolation algorithms. An interesting complication comes from the fact that the simulator provides results in the form of $97\times 97$ frames, which is not convenient for the standard structure of the autoencoders, since $97$ is not a number that can be achieved with the up-sampling layers. We tackle this with a zero padding that transforms the inputs to $128\times 128$ frames. Once again, we are concerned with \textit{extrapolation}, attempting to reconstruct the frames that correspond to the last $\lambda$ value, and use $5$ DOF. All the hyper-parameters of the compression and interpolation algorithms remain the same.

According to the results presented in Table \ref{table:precip_metrics_all}, the best performing combination is the CAE paired with the GP. Both the MAE and the MSE indicate significant improvement from the other compression and interpolation algorithms. The zero padding seems to affect the VAE negatively, which struggles to deal with the abrupt transitions between the zero-padded sections and the actual frames possibly due to the distributional aspect of the algorithm. 
The unusual structure of the data is not easy to capture as shown in Figure \ref{fig:precip_CAE}, where all the interpolation algorithms are shown for the CAE case. It is hard to distinguish anything in the earlier frames, mainly due to the common scaling and the fact that the DGP seems to be deviating significantly. In the frames that correspond to late times, we can observe the GP-reconstructed frames having some minor deviations from their true counterparts, but generally being able to capture the correct space of the circular features, while the other two methods deviate significantly.

Finally, in Figure \ref{fig:precip_CAE_Resids} we do not observe any specific pattern in the errors, which is expected from the structure of the data, but we can clearly see that the level of the errors is considerably lower for the late time frames in the GP case when compared to the other two.

\section{Conclusion}\label{sec:concl}
In this study, we used the ROM interpolation framework from our previous paper \cite{maulik2020latent} with the main purpose of demonstrating its use for emulating results for parameter values where data are not available. The effectiveness of this method is presented with three multiphase flow applications and results are compared with two other interpolation techniques; the DGPs, which are complex algorithms that use the GPs as building blocks, and the LSTMs, which have been used recently in literature \cite{maulik2021reduced} and have been found to work well for a class of flow problems. The results show that the choice of the best type of autoencoder is problem dependent, although CAE has a slight edge and is more versatile. In terms of the interpolation algorithms the GP and LSTM provided good results and the choice between the two is data-dependant. We believe that the main reason for the under-performance of the DGPs was over-parameterisation, with respect to the complexity of the generated flow patterns. In future work, we plan on replicating the same comparison for larger and more complex data-sets, in order to address this issue, though we expect to be confronted with new challenges, such as the choice of hyper-parameters, especially for the LSTM and DGP.

\begin{Backmatter}

\paragraph{Acknowledgments}
We thank Romit Maulik and Nesar Ramachandra from Argonne National Laboratory for their collaboration in a previous project that was used as a basis for this paper.

\paragraph{Funding Statement}
We acknowledge funding from the Engineering and Physical Sciences Research Council, UK, through the Programme Grant PREMIERE (EP/T000414/1), as well as funding through the Wave 1 of The UKRI Strategic Priorities Fund under the EPSRC Grant EP/T001569/1, particularly the \emph{Digital Twins for Complex Engineering Systems} theme within that grant, and the Royal Academy of Engineering through their support of OKM's PETRONAS/RAEng Research Chair in Multiphase Fluid Dynamics. IP acknowledges funding from the NUAcT fellowship scheme at Newcastle University and Imperial College Research Fellowship scheme at Imperial College London.

\paragraph{Competing Interests}
None

\paragraph{Data Availability Statement}
Replication data and code can be found in the Github repository: \url{https://github.com/themisbo/ROM_applications.git}.

\paragraph{Ethical Standards}
The research meets all ethical guidelines, including adherence to the legal requirements of the study country.

\paragraph{Author Contributions}
Conceptualisation, I.P., T.B.; Data curation, L.R.M.; Formal analysis, T.B., I.P.; Funding acquisition, O.K.M., I.P.; Investigation, T.B.; Methodology, I.P., T.B., L.R.M.; Project
administration, I.P., O.K.M.; Software, L.R.M., T.B.; Supervision, I.P., O.K.M.; Visualisation, T.B.; Writing—original
draft, T.B.; Writing—review and editing, I.P., L.R.M., O.K.M. All authors approved the final submitted draft.


\bibliographystyle{apalike}
\bibliography{bibliography}

\begin{thebibliography}{}

\bibitem[Abadi et~al., 2016]{abadi2016tensorflow}
Abadi, M., Agarwal, A., Barham, P., Brevdo, E., Chen, Z., Citro, C., Corrado,
  G.~S., Davis, A., Dean, J., Devin, M., et~al. (2016).
\newblock Tensorflow: Large-scale machine learning on heterogeneous distributed
  systems.
\newblock {\em arXiv preprint arXiv:1603.04467}.

\bibitem[Bar-Sinai et~al., 2019]{bar2019learning}
Bar-Sinai, Y., Hoyer, S., Hickey, J., and Brenner, M.~P. (2019).
\newblock Learning data-driven discretizations for partial differential
  equations.
\newblock {\em Proceedings of the National Academy of Sciences},
  116(31):15344--15349.

\bibitem[Berkooz et~al., 1993]{berkooz1993proper}
Berkooz, G., Holmes, P., and Lumley, J.~L. (1993).
\newblock The proper orthogonal decomposition in the analysis of turbulent
  flows.
\newblock {\em Annual review of fluid mechanics}, 25(1):539--575.

\bibitem[Bezanson et~al., 2017]{bezanson2017julia}
Bezanson, J., Edelman, A., Karpinski, S., and Shah, V.~B. (2017).
\newblock Julia: A fresh approach to numerical computing.
\newblock {\em SIAM review}, 59(1):65--98.

\bibitem[Bishop, 2006]{bishop2006pattern}
Bishop, C.~M. (2006).
\newblock {\em Pattern recognition and machine learning}.
\newblock Springer.

\bibitem[Damianou and Lawrence, 2013]{damianou2013deep}
Damianou, A. and Lawrence, N.~D. (2013).
\newblock Deep gaussian processes.
\newblock In {\em Artificial intelligence and statistics}, pages 207--215.
  PMLR.

\bibitem[Gardner et~al., 2018]{gardner2018gpytorch}
Gardner, J.~R., Pleiss, G., Bindel, D., Weinberger, K.~Q., and Wilson, A.~G.
  (2018).
\newblock Gpytorch: Blackbox matrix-matrix gaussian process inference with gpu
  acceleration.
\newblock In {\em Advances in Neural Information Processing Systems}.

\bibitem[Hochreiter and Schmidhuber, 1997]{hochreiter1997long}
Hochreiter, S. and Schmidhuber, J. (1997).
\newblock Long short-term memory.
\newblock {\em Neural computation}, 9(8):1735--1780.

\bibitem[Inguva et~al., 2020]{inguva2020numerical}
Inguva, P.~K., Mason, L.~R., Pan, I., Hengardi, M., and Matar, O.~K. (2020).
\newblock Numerical simulation, clustering, and prediction of multicomponent
  polymer precipitation.
\newblock {\em Data-Centric Engineering}, 1.

\bibitem[Inguva et~al., 2021]{inguva2021softmatter}
Inguva, P.~K., Walker, P.~J., Yew, H.~W., Zhu, K., Haslam, A.~J., and Matar,
  O.~K. (2021).
\newblock {Continuum-scale modelling of polymer blends using the Cahn-Hilliard
  equation: transport and thermodynamics}.
\newblock {\em Soft Matter}, 17:5645--5665.

\bibitem[Kim et~al., 2019]{kim2019deep}
Kim, B., Azevedo, V.~C., Thuerey, N., Kim, T., Gross, M., and Solenthaler, B.
  (2019).
\newblock Deep fluids: A generative network for parameterized fluid
  simulations.
\newblock In {\em Computer Graphics Forum}, volume~38, pages 59--70. Wiley
  Online Library.

\bibitem[Kingma and Welling, 2013]{kingma2013auto}
Kingma, D.~P. and Welling, M. (2013).
\newblock Auto-encoding variational bayes.
\newblock {\em arXiv preprint arXiv:1312.6114}.

\bibitem[LeCun et~al., 1995]{lecun1995convolutional}
LeCun, Y., Bengio, Y., et~al. (1995).
\newblock Convolutional networks for images, speech, and time series.
\newblock {\em The handbook of brain theory and neural networks},
  3361(10):1995.

\bibitem[Maulik et~al., 2021a]{maulik2020latent}
Maulik, R., Botsas, T., Ramachandra, N., Mason, L.~R., and Pan, I. (2021a).
\newblock Latent-space time evolution of non-intrusive reduced-order models
  using gaussian process emulation.
\newblock {\em Physica D: Nonlinear Phenomena}, 416:132797.

\bibitem[Maulik et~al., 2021b]{maulik2021reduced}
Maulik, R., Lusch, B., and Balaprakash, P. (2021b).
\newblock Reduced-order modeling of advection-dominated systems with recurrent
  neural networks and convolutional autoencoders.
\newblock {\em Physics of Fluids}, 33(3):037106.

\bibitem[Rasmussen and Williams, 2006]{williams2006gaussian}
Rasmussen, C.~E. and Williams, C. K.~I. (2006).
\newblock {\em Gaussian processes for machine learning}.
\newblock MIT Press Cambridge, MA.

\bibitem[Rohlfs et~al., 2018]{rohlfs2018wavemaker}
Rohlfs, W., Rietz, M., and Scheid, B. (2018).
\newblock Wavemaker: The three-dimensional wave simulation tool for falling
  liquid films.
\newblock {\em SoftwareX}, 7:211--216.

\bibitem[Salimbeni and Deisenroth, 2017]{salimbeni2017doubly}
Salimbeni, H. and Deisenroth, M. (2017).
\newblock Doubly stochastic variational inference for deep gaussian processes.
\newblock In {\em Advances in Neural Information Processing Systems}.

\bibitem[Scheid et~al., 2006]{scheid2006wave}
Scheid, B., Ruyer-Quil, C., and Manneville, P. (2006).
\newblock Wave patterns in film flows: modelling and three-dimensional waves.
\newblock {\em Journal of Fluid Mechanics}, 562:183--222.

\bibitem[Williams and Rasmussen, 1996]{williams1996gaussian}
Williams, C.~K. and Rasmussen, C.~E. (1996).
\newblock Gaussian processes for regression.
\newblock In {\em Adv. Neur. In.}, pages 514--520.

\bibitem[Zhuang et~al., 2020]{zhuang2020learned}
Zhuang, J., Kochkov, D., Bar-Sinai, Y., Brenner, M.~P., and Hoyer, S. (2020).
\newblock Learned discretizations for passive scalar advection in a 2-d
  turbulent flow.
\newblock {\em arXiv preprint arXiv:2004.05477}.

\end{thebibliography}

\end{Backmatter}

\end{document}